\documentclass[aps,twocolumn,pra,groupedaddress,superscriptaddress,amsmath,amssymb,amsthm,showpacs]{revtex4-1}
\usepackage{subfigure,hyperref,bbm,times}
\usepackage[T1]{fontenc}
\usepackage{braket}
\usepackage{amsthm}
\usepackage{epsfig}
\usepackage{color}
\usepackage{graphicx}
\usepackage{dcolumn}
\usepackage{bm}

\newcommand{\scal}[2]{\langle#1|#2\rangle}
\providecommand{\openone}{\leavevmode\hbox{\small1\kern-3.8pt\normalsize1}}

\begin{document}

\title{$N$ identical particles and one particle to entangle them all}

\author{Bruno Bellomo} 
\affiliation{Institut UTINAM - UMR 6213, CNRS, Observatoire des Sciences de
l'Univers THETA,  Universit\'{e} Bourgogne Franche-Comt\'{e}, F-25000 Besan\c con,
France}
\author{Rosario Lo Franco}
\email{rosario.lofranco@unipa.it}
\affiliation{Dipartimento di Energia, Ingegneria dell'Informazione e Modelli Matematici, Universit\`{a} di Palermo, Viale delle Scienze, Edificio 9, 90128 Palermo, Italy}
\affiliation{Dipartimento di Fisica e Chimica, Universit\`a di Palermo, via Archirafi 36, 90123 Palermo,
Italy}
\author{Giuseppe Compagno}
\affiliation{Dipartimento di Fisica e Chimica, Universit\`a di Palermo, via Archirafi 36, 90123 Palermo,
Italy}

\date{\today }


\begin{abstract}
In quantum information W states are a central class of multipartite entangled states because of their robustness against noise and use in many quantum processes.  
Their generation however remains a demanding task whose difficulty increases with the number of particles. 
We report a simple scalable conceptual scheme where a single particle in an ancilla mode works as entanglement catalyst of W state for other $N$ separated identical particles. A crucial novel aspect of the scheme, which exploits basically spatial indistinguishability, is its universality, being applicable without essential changes to both bosons and fermions. 
Our proposal represents a new paradigm within experimental preparation of many-particle entanglement based on quantum indistinguishability.
\end{abstract}

\maketitle

\section{Introduction}

Quantum entanglement represents nonclassical correlations among constituents of composite systems which make them intertwined independently of how far they are each other \cite{BrunnerReview}. It is well established that entanglement is essential as a resource by local operations and classical communication (LOCC) for implementing quantum information, computation and communication \cite{horodecki2009quantum,vedralReview}. Generation and control of entanglement in many-particle networks is thus very important from both theoretical and practical perspectives. 
A peculiar aspect is that there are states, like GHZ \cite{eisertReview}, W \cite{eisertReview,bengtssonChapter}, cluster \cite{PhysRevLett.86.910} and Dicke \cite{Dicke1954}, which belong to inequivalent classes of multipartite entanglement because they cannot be transformed into each other by LOCC \cite{ciracPRA2000,noriPRA2016}. 
Despite the exhaustive knowledge about bipartite entanglement, creation and characterization of multipartite entanglement remain challenging and debated \cite{eisertReview,bengtssonChapter,PhysRevLett.116.070504}.

Intense study has then focused on understanding the role as a resource in a given process of the different classes of multipartite entanglement \cite{eisertReview}. In this context, W states emerge as a particularly important class. Their entanglement is maximally robust against both noise and particle loss \cite{PhysRevLett.86.910,durPRA2001}, which makes nonclassical effects stronger for W states than for GHZ states for large number of particles \cite{PhysRevA.68.062306}. Furthermore, W states are central in quantum computation \cite{computeW2006}, secure quantum communication \cite{PhysRevA.60.2737,wangWsecure,jooWsecure,dongWsecure,caoWsecure}, teleportation \cite{caoWsecure,Gorbachev2003267,jooWteleport}, quantum heat engines \cite{dagEntropy} and quantum key distribution \cite{sharmaJPB2008}. Designing \cite{sharmaJPB2008,liu2014JAP,yuPRA2007,zouPRA2002,PhysRevA.66.064301,kangSciRep2016,biswasJMP,swekePRA,gaoPRA,Moreno2016,krenn2017,gaoxiangPRA,He2014QIP,Yesilyurt:16,PhysRevA.67.034302,PhysRevA.67.022302,PhysRevA.71.052310, songPRA2007,zhangPRA2006,dengPRA2006,ikuta2011,imotoNJP,perezPRA,han2015SciRep,zang2015SciRep,Zang:16} and realizing \cite{grafeNatPhoton,tashima2009,tashimaPRL,blattNature2005,4opticalmodes2009,neeleyNature2010,choiNature2010,altomareNatPhys2010} production schemes of this class of multipartite states has thus attracted great attention.

The many theoretical proposals for generating W states work for specific systems and require in general precise control of interparticle and particle-environment interactions, nonlocal external operations, initially entangled photon pairs, fusion of previously created W states with ancilla photons and complex network gates \cite{sharmaJPB2008,liu2014JAP,yuPRA2007,zouPRA2002,PhysRevA.66.064301,kangSciRep2016,biswasJMP,swekePRA,gaoPRA,Moreno2016,krenn2017,gaoxiangPRA,He2014QIP,Yesilyurt:16,PhysRevA.67.034302,PhysRevA.67.022302,PhysRevA.71.052310, songPRA2007,zhangPRA2006,dengPRA2006,ikuta2011,imotoNJP,perezPRA,han2015SciRep,zang2015SciRep,Zang:16}. Combinations of these requisites make the implementation very demanding. So far the W states with the largest number of particles observed in the laboratory consist of eight two-level trapped ions \cite{blattNature2005}, while for instance this number is lowered to four particles for polarized photons \cite{tashimaPRL}. A key step towards simpler reliable generation protocols is to find conceptual schemes based on fundamental mechanisms valid for general systems. Since distributed quantum networks are typically made of identical particles (e.g., electrons, atoms, photons, nuclei, quantum dots), a natural candidate to act as basic entangling resource is quantum indistinguishability of the particles themselves \cite{krenn2017,lofranco2015quantum,sciaraSchmidt,tichyPRA,plenio2014PRL}.   
 
In this work we introduce a universal conceptual scheme, valid for both bosons and fermions, which creates a W state of $N$ separated identical particles by exploiting only their spatial indistinguishability and random destination sources. The key ingredient is supplied by a single particle staying in an ancilla spatial mode which, after postselection at a given step of the protocol, serves as an entanglement catalyst for other $N$ particles. The number of circuital elements scales linearly with the number of particles, which are initially independent and uncorrelated. Simplicity and generality of the scheme with respect to previous proposals based on indistinguishability make it a novel promising blueprint for experimental generation of many-particle entanglement in different contexts, from quantum optics to solid state and condensed matter. 

Following a recent nonstandard particle-based approach to treat identical particles without using labels \cite{lofranco2015quantum}, we indicate an elementary pure state of a $N$-particle composite system by $\ket{\varphi_1,\varphi_2,\ldots,\varphi_N}$, which represents a particle in the state $\varphi_1$, a particle in $\varphi_2$ and so on. Such a state is in general an indivisible object whose normalization constant is to be determined by single-particle probability amplitudes (see Appendix A). Each single-particle state is characterized by a spatial mode, indicated with a capital letter (e.g., M), and a given pseudospin $\sigma$, whose basis in a given direction is denoted by $\{\ket{\uparrow},\ket{\downarrow}\}$. 
An aspect of this formalism is that, when each particle of a subsystem is spatially separated from the particles in the other subsystems and only under local measurements, the overall elementary state of indistinguishable particles can be written as a tensor (separable) product of subsystem states \cite{lofranco2015quantum}. Under these assumptions the cluster decomposition principle, stating that distant experiments yield independent results, holds \cite{peresbook} and the identical particles behave like distinguishable individually addressable ones.

\begin{figure}
  \centering
  \includegraphics[width=0.49 \textwidth]{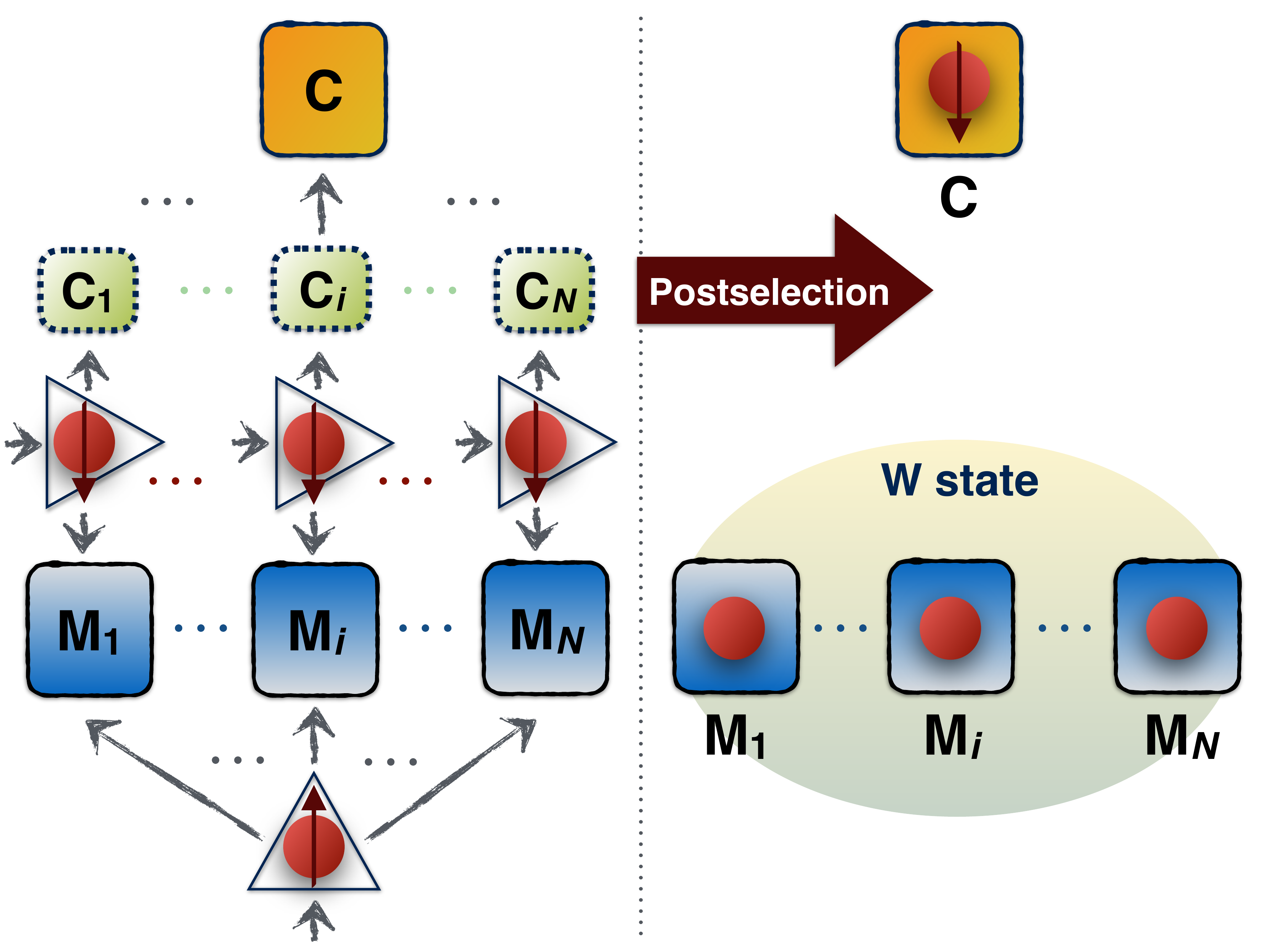}
  \caption{\textbf{Ancilla mode-based scheme with $N+1$ identical particles.} $N$ identical particles with pseudospin $\downarrow$ are equally split in two separated modes while the one with pseudospin $\uparrow$ is equally split into the modes $\mathrm{M}_i$ ($i=1,\ldots,N$). For convenience, in the left panel the particles are placed at the network nodes (triangles). After postselection and reaching $\mathrm{C}$, a W state is generated among the pseudospins of $N$ particles in modes $\mathrm{M}_1,\ldots,\mathrm{M}_N$.}
  \label{figN+1part}
\end{figure}

\section{$N$-particle W state generation}

We take a system of $N+1$ indistinguishable particles, initially uncorrelated and spatially separated, where $N$ particles have pseudospin $\downarrow$ and one pseudospin $\uparrow$. The overall initial (normalized) state is $\ket{\Phi^{(N+1)}_0}=\ket{\mathrm{A_1}\downarrow,\mathrm{A_2}\downarrow,\ldots,\mathrm{A}_{N+1}\uparrow}$. Each particle then goes to a network node, as illustrated in Fig.~\ref{figN+1part}, after which the (normalized) global state is
\begin{equation}\label{N+1partstate}
\ket{\Phi^{(N+1)}}=\ket{\varphi_1,\varphi_2, \ldots,  \varphi_{N+1}},
\end{equation}
where the $N$ particles in $\downarrow$ and the $N+1$-th particle in $\uparrow$ are transformed, respectively, as
\begin{eqnarray}\label{N+1singlepartstates}
\ket{\varphi_i}&=& (\ket{\mathrm{M}_i\downarrow}+
\ket{\mathrm{C}_i\downarrow})/\sqrt{2},\quad (i=1,2,\ldots,N) \nonumber\\
\ket{\varphi_{N+1}}&=& \sum_{i=1}^ {N}\ket{\mathrm{M}_i\uparrow}/\sqrt{N}.
\end{eqnarray}
The nodes of the network preparing these orthonormal one-particle states are random destination sources behaving like beam-splitters \cite{sciarrinoPRA} (we shall later discuss the experimental implementation). 
The linear expansion (see Appendix A) of $\ket{\Phi^{(N+1)}}$ according to the explicit expressions of Eq.~(\ref{N+1singlepartstates}) thus contains $N \times 2^N$ components. At this stage we perform a postselection on modes $\mathrm{M}_i$ such that each of these modes contains one particle alone. This operation produces the state $\ket{\Phi_p^{(N+1)}} = (\ket{\mathrm{C}_1\downarrow,\mathrm{M}_2\downarrow,\ldots,\mathrm{M}_N\downarrow,\mathrm{M}_1\uparrow}+\ldots+\ket{\mathrm{M}_1\downarrow,\mathrm{M}_2\downarrow,\ldots,\mathrm{C}_N\downarrow,\mathrm{M}_N\uparrow})/\sqrt{N}$. This state is obtained from the global state with probability $\mathcal{P}=|\scal{\Phi^{(N+1)}_p}{\Phi^{(N+1)}}|^2=1/2^N$ (see Appendix B). The particle present in the $i$-th intermediate mode $\mathrm{C}_i$ ($i=1,\ldots, N$) is successively sent to a common ancilla mode $\mathrm{C}$, which deterministically leads to the final state 
\begin{eqnarray}\label{N+1partstate projected}
\ket{\Psi^{(N+1)}} &=& (1/\sqrt{N}) (\ket{\mathrm{C}\downarrow,\mathrm{M}_2\downarrow,\ldots,\mathrm{M}_N\downarrow,\mathrm{M}_1\uparrow}\nonumber\\
&+&\ket{\mathrm{M}_1\downarrow,\mathrm{C}\downarrow,\ldots,\mathrm{M}_N\downarrow,\mathrm{M}_2\uparrow}+\ldots \nonumber\\
&+&\ket{\mathrm{M}_1\downarrow,\mathrm{M}_2\downarrow,\ldots,\mathrm{C}\downarrow,\mathrm{M}_N\uparrow}).
\end{eqnarray}
Using the symmetrization rule with respect to the swapping of single-particle state positions (see Eq.~(\ref{Nsimmetry}) of Appendix A), taking into account that the particle in mode $\mathrm{C}$ is separated from the other particles and assuming LOCC from now on, the state of Eq.~(\ref{N+1partstate projected}) can be written as a tensor product of an $N$-particle state and a single-particle state as 
\begin{equation}\label{Psi}
\ket{\Psi^{(N+1)}} \equiv \eta \ket{\mathrm{W}_N} \otimes \ket{\mathrm{C} \downarrow},
\end{equation} 
where $\eta=\pm1$ for bosons and fermions, respectively, and
\begin{eqnarray}\label{WN}
\ket{\mathrm{W}_N}&  =&(\ket{\mathrm{M_1} \uparrow, \mathrm{M_2}\downarrow, \mathrm{M_3}\downarrow, \ldots , \mathrm{M}_N\downarrow}\nonumber\\
&+& \ket{\mathrm{M_1} \downarrow, \mathrm{M_2}\uparrow, \mathrm{M_3}\downarrow,\ldots ,  \mathrm{M}_N\downarrow} + \ldots \nonumber\\
&+& \ket{\mathrm{M_1} \downarrow, \mathrm{M_2}\downarrow, \mathrm{M_3}\downarrow,\ldots, \mathrm{M}_N\uparrow})/\sqrt{N}.
\end{eqnarray}
The protocol therefore creates a $N$-particle W state $\ket{\mathrm{W}_N}$ among the pseudospins of $N$ particles in separated modes $\mathrm{M}_1,\ldots,\mathrm{M}_N$, which is a superposition of states such that $N-1$ particles have pseudospin $\ket{\downarrow}$ and one has $\ket{\uparrow}$ \cite{eisertReview}. Within the resource theory of LOCC, $\ket{\mathrm{W}_N}$ represents a genuine multipartite entangled state among $N$ individually addressable identical particles.

\begin{figure}[t]
  \centering
  \includegraphics[width=0.45 \textwidth]{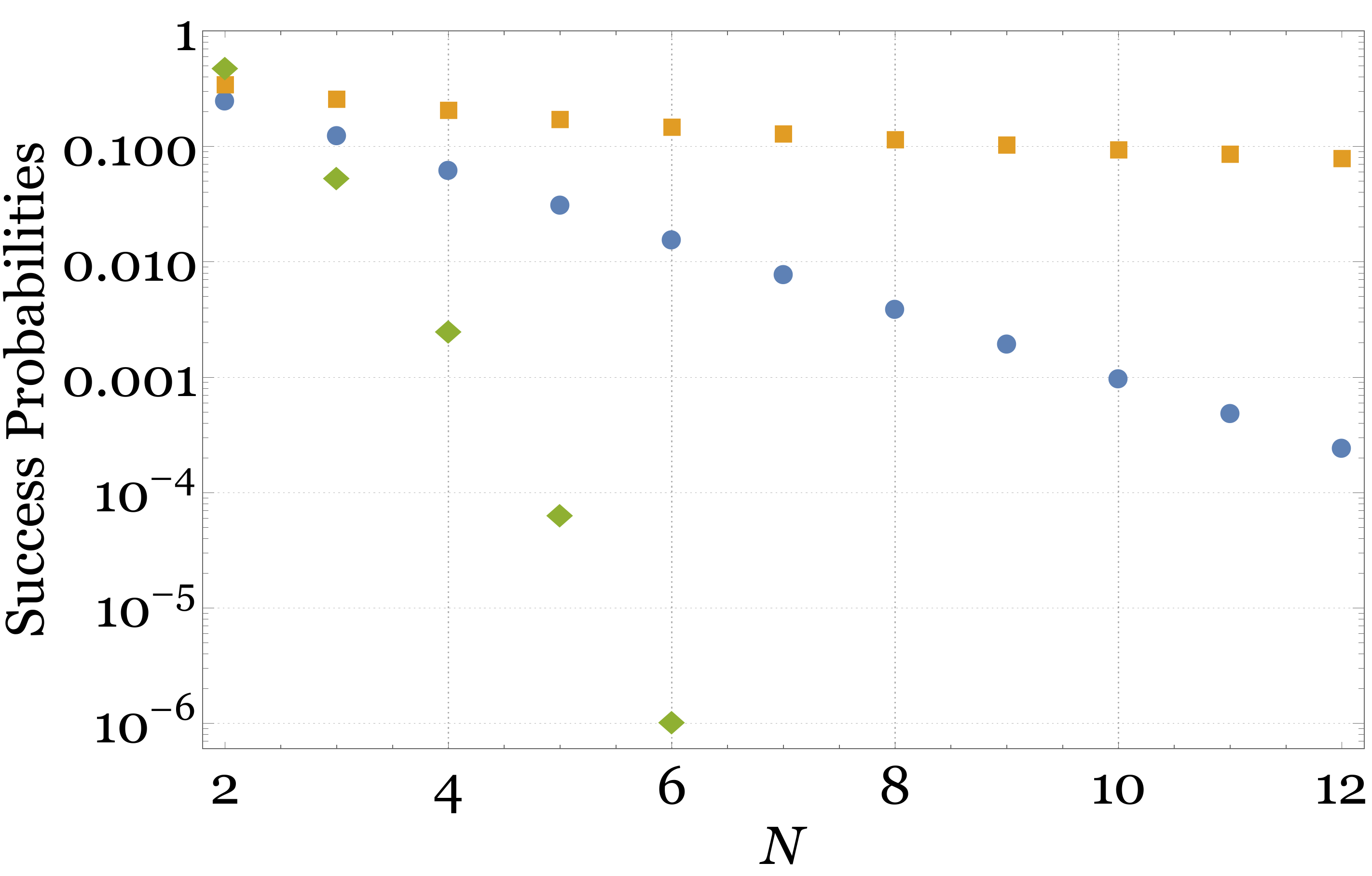}
  \caption{\textbf{Success probabilities.} Plots of the probabilities of success to create a $N$-particle W state by the ancilla mode-based scheme, for bosons $\mathcal{P}_\mathrm{b}=1/2^N$ (blue points) and fermions $\mathcal{P}_\mathrm{f}=1/(N+1)$ (orange squares), and by the extraction protocol for bosons $\widetilde{\mathcal{P}}_\mathrm{b} =1/[N^{(N-1)} (N-1)!]$ (green diamonds).}\label{figcomparison}
\end{figure}

The scheme above equally works for both bosons and fermions. However, we point out that for fermions the state $\ket{\Psi^{(N+1)}}$ of Eq.~(\ref{Psi}), and thus $\ket{\mathrm{W}_N}$, can be achieved more efficiently by performing the postselection on modes $\mathrm{M}_i$ after the particles are distributed among the common mode $\mathrm{C}$ and the modes $\mathrm{M}_i$, that means $\mathrm{C}_i=\mathrm{C}$ in Eq.~(\ref{N+1singlepartstates}). We indicate with $\ket{\Phi_f^{(N+1)}}$ the state of Eq.~(\ref{N+1partstate}) under this condition. In the fermion linear expansion of $\ket{\Phi_f^{(N+1)}}$, all the terms where mode $\mathrm{C}$ would appear more than one time are forbidden by the Pauli exclusion principle. As a consequence, the probability to obtain $\ket{\Psi^{(N+1)}}$ from the global state is now (see Appendix B) $\mathcal{P}_\mathrm{f}=|\scal{\Psi^{(N+1)}}{\Phi_f^{(N+1)}}|^2=1/(N+1)$ (linear scaling) against $\mathcal{P}_\mathrm{b}=\mathcal{P}=1/2^N$ (exponential scaling) for bosons.
We also notice that, in the case of bosons, performing postselection when particles are distributed among the common ancilla mode $\mathrm{C}$ and the modes $\mathrm{M}_i$, would decrease the success probability to $1/[\sum_{m=0}^N N!/(N-m)!]$, which decays much faster than $1/2^N$ (see Appendix B). The probabilities of success to produce $\ket{\mathrm{W}_N}$ are plotted in Fig.~\ref{figcomparison}, where the greater efficiency of the scheme for fermions than for bosons is evident for large $N$. The proposed scheme proves that, after postselection, a single particle in an ancilla mode $C$ is capable thanks to indistinguishability to act as an entanglement catalyst among the pseudospins of the $N$ remaining identical particles in the separated spatial modes.  

\section{Comparison with an extraction-based scheme} 

We now compare the performance of the above scheme with a generation protocol which we would obtain by a generalization of the entanglement extraction from identical particles \cite{plenio2014PRL}. The latter procedure indeed constitutes a natural strategy to exploit entanglement among identical particles due to indistinguishability. 
The initial configuration is made of $N$ particles in the same mode $\mathrm{M}$, all having pseudospin $\downarrow$ but one with pseudospin $\uparrow$, that is $\ket{\mathrm{M}\uparrow, \mathrm{M}\downarrow,\ldots,\mathrm{M}\downarrow}/\sqrt{(N-1)!}$. Notice that this state is allowed only for bosons. Each particle then tunnels towards $N$ separated modes $\mathrm{M}_i$ ($i=1,\ldots,N$) with the same probability amplitude maintaining the pseudospin state $\sigma$:
$\ket{\mathrm{M} \sigma}\rightarrow\sum_{i} (1/\sqrt{N})\ket{\mathrm{M}_i\sigma}$. This leads to the extracted global state $\ket{\Psi^{(N)}}=\ket{\mathrm{M}_\mathrm{nl}\uparrow,\mathrm{M}_\mathrm{nl}\downarrow,\ldots,\mathrm{M}_\mathrm{nl}\downarrow}/\sqrt{(N-1)!}$, where $\ket{\mathrm{M}_\mathrm{nl}}\equiv \sum_{i} (1/\sqrt{N})\ket{\mathrm{M}_i}$ represents the common nonlocal mode for the identical particles \footnote{We stress that also this output state is forbidden for fermions, so that the protocol does not work for this kind of particles even if they initially come from separated spatial modes rather than from the same mode.}. It is straightforward to see that, by projecting the global state $\ket{\Psi^{(N)}}$ onto the subspace such that each spatial mode contains one particle alone, we are left with the W state $\ket{\mathrm{W}_N}$ of Eq.~(\ref{WN}) with probability $\widetilde{\mathcal{P}}_\mathrm{b}=|\scal{\mathrm{W}_N}{\Psi^{(N)}}|^2=1/[N^{(N-1)} (N-1)!]$ (see Appendix B). 
In Fig.~\ref{figcomparison} this probability is compared to the success probabilities of the ancilla mode-based scheme for bosons and fermions. The extraction protocol results significantly disadvantageous already for $N\geq3$. Moreover, the introduction of the ancilla mode $C$ (see Fig.~\ref{figN+1part}) greatly simplifies the implementation of the proposed scheme with respect to the extraction one, since only one particle (the $N+1$-th) must be sent to $N$ spatial modes.

We now compare the most recent proposals for W state generation with ours. Most of the former are based on fusion of preexisting W states that requires precise control of interactions and sequences of nonlocal gates \cite{zang2015SciRep,han2015SciRep,Zang:16,Yesilyurt:16}: albeit some of these schemes are in principle deterministic \cite{Zang:16,Yesilyurt:16}, these requisites make their realization challenging. Other schemes utilizing path identity are given for a small number of particles ($N = 4$) \cite{krenn2017}. Finally, all of these proposals are devised for particles of a given kind, such as photons for example \cite{Yesilyurt:16,krenn2017}. Our method, besides its simplicity and straightforward scalability, appears to be the only one applicable to any type of particles (bosons or fermions).

\section{Discussion on the experimental feasibility}

The conceptual scheme of Fig.~\ref{figN+1part} can be implemented in different experimental contexts by currently available technologies. Its first possible realization is by polarized photons in quantum optical setups, where horizontal and vertical polarizations encode the two pseudospin states. The initial independent photons can be generated either by standard single-photon sources \cite{zeilingerPRA,somaschiNatPhoton} or by coherently harnessing a single photon of a polarization entangled pair \cite{sciarrino2017}, while the random destination sources at the nodes of the network are given by beam splitters \cite{sciarrinoPRA}. The photon which has to be equally split into the $N$ modes $\mathrm{M}_i$ can travel along a path with $N-1$ cascaded beam splitters (notice that for odd $N$ the first beam splitter of the cascade must be unbalanced). The required postselection can be performed by single-photon quantum non-demolition detectors placed in each mode $\mathrm{M}_i$ \cite{QND1,QND3,QND4}. Finally, the optical paths of the particles traveling to the common ancilla mode $\mathrm{C}$ must, as usual, be adjusted in order to maintain indistinguishability and interference effects \cite{Ra2013}. Such a linear optical setup is expected to be exploitable for simply generating the first W state with a number of photons larger than four, which is the current achievement for polarized photons \cite{tashimaPRL}. 
Due to the existing toolkit for linear-optics quantum computing in circuit quantum electrodynamics \cite{PhysRevLett.104.230502,QND2}, our conceptual scheme is also amenable to be realized in the solid state with superconducting flux qubits, involving both bosons and fermions \cite{solano2017}. 
Another possible field of application is in condensed matter where quantum degenerate bosons or fermions can be prepared in independent sites of a lattice and then suitably harnessed \cite{PhysRevLett.117.213001}. In this context, the particles can tunnel from their initial site to other ones with probability amplitudes adjustable by varying external parameters such as gate voltages, magnetic fields and laser beams, thus creating the analogous of beam splitter operations \cite{plenio2014PRL}. 
Looking at the success probabilities of our ancilla mode-based scheme, it also appears feasible the reaching of a W state with $N>8$, that is the current general limit obtained with trapped ions \cite{blattNature2005}. 
The scheme is in fact scalable, being straightforwardly generalizable to any number of particles, as shown in Fig.~\ref{figN+1part}, with success probabilities which remain significantly larger than zero for values of $N$ ordinarily achievable in experimental contexts. For instance, taking as a threshold the success probability $P_\mathrm{exp}\sim 10^{-9}$ of the most recent experiment for the production of ten-photon GHZ states \cite{exptenphoton}, our procedure is in principle capable to create, aside from experimental uncertainties, a W state with $N\sim 30$ bosons and $N\sim10^9$ fermions.

\section{Conclusions}

We have proposed a scheme to generate $N$-particle entangled states of the W class which is scalable and universal, working for both bosons and fermions. The crucial ingredient for the working of this scheme is the introduction of a common ancilla mode which, by exploiting the spatial indistinguishability of identical particles, postselection and only local operations, enables the particle being there to entangle the other $N$ particles. The importance of the common ancilla mode is evinced from the fact that tracing out modes $\mathrm{C}_i$ at the intermediate stage of the scheme would give a maximally mixed state in modes $\mathrm{M}_i$. 

We stress that the proposed scheme does not work if nonidentical particles are employed. To understand this point it is sufficient to consider the case $N=2$ of the scheme, that requires three particles. Starting from the initial state $\ket{\Phi^{(3)}_{abc}}=\ket{\varphi_1^a,\varphi_2^b,\varphi_3^c}$ of Eq.~(\ref{N+1partstate}), where the particles are now labeled $a$, $b$, $c$ in order to be distinguishable from one another, at the end of the protocol one gets the corresponding state of Eq.~(\ref{N+1partstate projected}) with labels: $\ket{\Psi^{abc}} = (\ket{\mathrm{C}\downarrow_a, \mathrm{M_2} \downarrow_b,\mathrm{M_1}\uparrow_c}+\ket{\mathrm{M_1} \downarrow_a, \mathrm{C}\downarrow_b,  \mathrm{M_2}\uparrow_c})/\sqrt{2}$. In such a state a given particle does not have an assigned localized mode. As a first consequence, the mode $\mathrm{C}$ cannot be isolated; moreover, a local measurement of the pseudospin of particle $a$ on $\mathrm{M}_1$ always renders the outcome $\ket{\downarrow_a}$, leaving particles $b$ and $c$ respectively in  $\ket{\mathrm{C}\downarrow_b}$ and $\ket{\mathrm{M_2}\uparrow_c}$ without any correlation between pseudospins in separated modes. These arguments hold for any number of particles. 

We remark that the use of the recent non-standard approach to identical particles \cite{lofranco2015quantum} plays a crucial role in maintaining our analysis very simple and obtaining the results in a straightforward way. Compared to previous proposals, our scheme constitutes a good compromise between success probability and simplicity of the network, being more efficient for fermions than for bosons. We have discussed its experimental implementation showing that it is feasible by current technologies in different contexts. Universality, scalability and simplicity make the proposed scheme a novel experimentally realizable paradigm within the generation of multiparticle entanglement based on quantum indistinguishability.

\appendix

\section{Probability amplitude between $N$-particle states, linearity and particle statistics}

The approach used to describe identical particle states is the particle-based one without labels, recently introduced for systems of two particles \cite{lofranco2015quantum,sciaraSchmidt}. This approach is straightforwardly generalizable to states of $N$ particles. While a complete treatment will be done elsewhere, here we simply give the core of the approach constituted by the definition of the probability amplitude and the rule it induces for one-particle states permutation depending on particle statistics (bosons or fermions). 

A pure state representing, respectively, a particle in the state $\varphi_1$, $\varphi_2$ and so on is $\ket{\varphi_1,\varphi_2,\ldots,\varphi_N}$. All the physical information of the system is contained in the probability amplitude expressed by the scalar product $\scal{\varphi'_1,\varphi'_2,\ldots,\varphi'_N}{\varphi_1,\varphi_2,\ldots,\varphi_N}$. Generalizing the two-particle probability amplitude \cite{lofranco2015quantum}, the $N$-particle probability amplitude is defined as
\begin{eqnarray}\label{Nscalarproduct}
&\scal{\varphi'_1,\varphi'_2,\ldots,\varphi'_N}{\varphi_1,\varphi_2,\ldots,\varphi_N}:=&\nonumber\\ 
&\sum_P \eta^{n_P} \scal{\varphi'_1}{\varphi_{P_1}}\scal{\varphi'_2}{\varphi_{P_2}}\cdots\scal{\varphi'_N}{\varphi_{P_N}},&
\end{eqnarray} 
where the sum is taken over all different permutations $P=\{P_1,P_2,\ldots,P_N\}$ acting on the $N$ particles in the ket state and $n_P$ is the number of transpositions in each permutation; $\eta=+1$ is for bosons and $\eta=-1$ for fermions. 

Linearity of the $N$-particle state vector with respect to each one-particle state immediately follows from the linearity of the one-particle amplitudes: for a one-particle state $\ket{\varphi_i}=a\ket{\varphi_a^{(i)}}+b\ket{\varphi_b^{(i)}}$ ($|a|^2+|b|^2=1$), one has $\ket{\varphi_1,\ldots,\varphi_i,\ldots,\varphi_N}=a\ket{\varphi_1,\ldots,\varphi_a^{(i)},\ldots,\varphi_N}+b\ket{\varphi_1,\ldots,\varphi_b^{(i)},\ldots,\varphi_N}$. 

The right-hand side of equation~(\ref{Nscalarproduct}) induces a symmetry with respect to the swapping of two one-particle state positions within the $N$-particle state vector. In fact, from the equality $\scal{\varphi'_1,\varphi'_2,\ldots,\varphi'_N}{\varphi_1,\ldots,\varphi_i,\ldots,\varphi_j,\ldots,\varphi_N}=\eta \scal{\varphi'_1,\varphi'_2,\ldots,\varphi'_N}{\varphi_1,\ldots,\varphi_j,\ldots,\varphi_i,\ldots,\varphi_N}$, one  obtains
\begin{eqnarray}\label{Nsimmetry}
&\ket{\varphi_1,\ldots,\varphi_i,\ldots,\varphi_j,\ldots,\varphi_N}=&\nonumber\\
&\eta\ket{\varphi_1,\ldots,\varphi_j,\ldots,\varphi_i,\ldots,\varphi_N},&
\end{eqnarray} 
for any $i,j=1,\ldots,N$. Thus, particle statistics is automatically encompassed in the approach \cite{lofranco2015quantum}.

The normalized state vector corresponding to a $N$-particle state is $\ket{\Phi^{(N)}}=(1/\mathcal{N})\ket{\varphi_1,\ldots,\varphi_N}$, such that $\scal{\Phi^{(N)}}{\Phi^{(N)}}=1$, with normalization constant $\mathcal{N}=\sqrt{\scal{\varphi_1,\ldots,\varphi_N}{\varphi_1,\ldots,\varphi_N}}$ obtained by application of equation~(\ref{Nscalarproduct}); for orthogonal one-particle states, $\scal{\varphi_i}{\varphi_j}=\delta_{ij}$, one has $\mathcal{N}=1$.

\section{Success probabilities}

In this section we give the explicit calculations to obtain the probabilities to generate the W state for bosons and fermions by our ancilla mode-based scheme and by the extraction-based scheme.

The probability to generate the desired W state of Eq.~(\ref{WN}) of the main text for bosons depends on the probability to obtain the intermediate projected state  
\begin{eqnarray}
\ket{\Phi_p^{(N+1)}} &=& \frac{1}{\sqrt{N}} (\ket{\mathrm{C}_1\downarrow,\mathrm{M}_2\downarrow,\ldots,\mathrm{M}_N\downarrow,\mathrm{M}_1\uparrow}\nonumber\\
&+&\ket{\mathrm{M}_1\downarrow,\mathrm{C}_2\downarrow,\ldots,\mathrm{M}_N\downarrow,\mathrm{M}_2\uparrow}+\ldots\nonumber\\
&+&\ket{\mathrm{M}_1\downarrow,\mathrm{M}_2\downarrow,\ldots,\mathrm{C}_N\downarrow,\mathrm{M}_N\uparrow}),
\end{eqnarray}
after postselection on the global state $\ket{\Phi^{(N+1)}}$ defined in Eqs.~(\ref{N+1partstate}) and (\ref{N+1singlepartstates}) of the main text. The probability is thus given by $\mathcal{P}_\mathrm{b}=|\scal{\Phi^{(N+1)}_p}{\Phi^{(N+1)}}|^2$. It is immediate to see that only $N$ terms give nonzero contributions (namely, equal to one) to the scalar product and, taking into account the normalization constants, one gets
\begin{equation}
\mathcal{P}_\mathrm{b}=|\scal{\Phi^{(N+1)}_p}{\Phi^{(N+1)}}|^2=\left(\frac{N}{\sqrt{N }\sqrt{N2^N}} \right)^2=\frac{1}{2^N}.
\end{equation}

The probability of success for fermions is given by the probability to get the final state $\ket{\Psi^{(N+1)}} = \eta \ket{\mathrm{W}_N} \otimes \ket{\mathrm{C} \downarrow}$, where $\eta=-1$ and $\ket{\mathrm{W}_N}$ is the desired W state of Eq.~(\ref{WN}) of the main text after postselection. The fermions are sent directly to the common mode $\mathrm{C}$ by the network nodes and, once reaching this mode, postselection is performed. The state $\ket{\Phi_f^{(N+1)}}$ for fermions, before postselection, is given by Eqs.~(\ref{N+1partstate}) and (\ref{N+1singlepartstates}) of the main text with $\mathrm{C}_i = \mathrm{C}$ ($i=1,2,\dots,N$). The important difference with the previous case for bosons is that now the terms appearing in the linear expansion of the global state $\ket{\Phi_f^{(N+1)}}$ before postselection must take into account the Pauli exclusion principle. The forbidden terms in its linear expansion, due to the Pauli exclusion principle, are therefore those where there are two or more particles in the same mode with the same pseudospin. This event can only happen for particles in the ancilla mode $\mathrm{C}$, all having pseudospin $\downarrow$, which leads to cancel out any term of the linear expansion where the mode $\mathrm{C}$ appears more than one time. The number of remaining orthonormal terms is in particular $N(N+1)$, so that the normalization constant of $\ket{\Phi_f^{(N+1)}}$ for fermions is $1/\sqrt{N(N+1)}$. The success probability is then $\mathcal{P}_\mathrm{f}=|\scal{\Psi^{(N+1)}}{\Phi_f^{(N+1)}}|^2$, where $\ket{\Psi^{(N+1)}}$ is defined in Eq. (\ref{Psi}) of the main text and, once again, only $N$ terms give nonzero contributions (namely, equal to one) in the scalar product. Taking into account the normalization constants, one finds
\begin{eqnarray}
\mathcal{P}_\mathrm{f}=|\scal{\Psi^{(N+1)}}{\Phi_f^{(N+1)}}|^2&=&\left(\frac{N}{\sqrt{N}\sqrt{N(N+1)}} \right)^2 \nonumber\\
&=&\frac{1}{N+1}.
\end{eqnarray}

If also for bosons the postselection is made after particles in modes 
C$_{i}$ arrive at the common mode C, one has the global state $\ket{\Phi_b^{(N+1)}}$ given by Eqs.~(\ref{N+1partstate}) and (\ref{N+1singlepartstates}) with $\mathrm{C}_i = \mathrm{C}$ ($i=1,2,\dots,N$). The success probability is given by $\bar{\mathcal{P}}_\mathrm{b}=|\scal{\Psi^{(N+1)}}{\Phi_b^{(N+1)}}|^2$. Differently from the above case of fermions, all the terms of the linear expansion of $\ket{\Phi_b^{(N+1)}}$ where the ancilla mode C appears more than one time are allowed. The normalization constant of the global state $\ket{\Phi_b^{(N+1)}}$ is now $1/\sqrt{N \sum_{m=0}^{N}[N!/(N-m)!]}$. Since only $N$ terms give nonzero contributions to the scalar product, the success probability is then obtained from the global state as
\begin{eqnarray}
\bar{\mathcal{P}}_\mathrm{b}&=&|\scal{\Psi^{(N+1)}}{\Phi_b^{(N+1)}}|^2 \nonumber \\ &=&\left(\frac{N}{\sqrt{N}\sqrt{N \sum_{m=0}^{N}[N!/(N-m)!]}} \right)^2 \nonumber\\
&=&\frac{1}{ \sum_{m=0}^{N}[N!/(N-m)!]}.
\end{eqnarray}
 
Finally, let us consider the protocol based on entanglement extraction from identical particles described in the main text as a comparison with our proposed scheme. By taking the explicit linear expansion of the extracted normalized global state $\ket{\Psi^{(N)}}=\ket{\mathrm{M}_\mathrm{nl}\uparrow,\mathrm{M}_\mathrm{nl}\downarrow,\ldots,\mathrm{M}_\mathrm{nl}\downarrow}/\sqrt{(N-1)!}$, where $\ket{\mathrm{M}_\mathrm{nl}}\equiv \sum_{i} (1/\sqrt{N})\ket{\mathrm{M}_i}$, the probability to obtain the W state $\ket{\mathrm{W}_N}$ of Eq.~(\ref{WN}) after postselection is given by
\begin{eqnarray}
\widetilde{\mathcal{P}}_\mathrm{b}=|\scal{\mathrm{W}_N}{\Psi^{(N)}}|^2&=&
\left(\frac{N}{\sqrt{N}\sqrt{N^N}\sqrt{(N-1)!}}\right)^2 \nonumber\\
&=&\frac{1}{(N-1)!\ N^{(N-1)}}.
\end{eqnarray}



\begin{thebibliography}{74}%
\makeatletter
\providecommand \@ifxundefined [1]{%
 \@ifx{#1\undefined}
}%
\providecommand \@ifnum [1]{%
 \ifnum #1\expandafter \@firstoftwo
 \else \expandafter \@secondoftwo
 \fi
}%
\providecommand \@ifx [1]{%
 \ifx #1\expandafter \@firstoftwo
 \else \expandafter \@secondoftwo
 \fi
}%
\providecommand \natexlab [1]{#1}%
\providecommand \enquote  [1]{``#1''}%
\providecommand \bibnamefont  [1]{#1}%
\providecommand \bibfnamefont [1]{#1}%
\providecommand \citenamefont [1]{#1}%
\providecommand \href@noop [0]{\@secondoftwo}%
\providecommand \href [0]{\begingroup \@sanitize@url \@href}%
\providecommand \@href[1]{\@@startlink{#1}\@@href}%
\providecommand \@@href[1]{\endgroup#1\@@endlink}%
\providecommand \@sanitize@url [0]{\catcode `\\12\catcode `\$12\catcode
  `\&12\catcode `\#12\catcode `\^12\catcode `\_12\catcode `\%12\relax}%
\providecommand \@@startlink[1]{}%
\providecommand \@@endlink[0]{}%
\providecommand \url  [0]{\begingroup\@sanitize@url \@url }%
\providecommand \@url [1]{\endgroup\@href {#1}{\urlprefix }}%
\providecommand \urlprefix  [0]{URL }%
\providecommand \Eprint [0]{\href }%
\providecommand \doibase [0]{http://dx.doi.org/}%
\providecommand \selectlanguage [0]{\@gobble}%
\providecommand \bibinfo  [0]{\@secondoftwo}%
\providecommand \bibfield  [0]{\@secondoftwo}%
\providecommand \translation [1]{[#1]}%
\providecommand \BibitemOpen [0]{}%
\providecommand \bibitemStop [0]{}%
\providecommand \bibitemNoStop [0]{.\EOS\space}%
\providecommand \EOS [0]{\spacefactor3000\relax}%
\providecommand \BibitemShut  [1]{\csname bibitem#1\endcsname}%
\let\auto@bib@innerbib\@empty
\bibitem [{\citenamefont {Brunner}\ \emph {et~al.}(2014)\citenamefont
  {Brunner}, \citenamefont {Cavalcanti}, \citenamefont {Pironio}, \citenamefont
  {Scarani},\ and\ \citenamefont {Wehner}}]{BrunnerReview}%
  \BibitemOpen
  \bibfield  {author} {\bibinfo {author} {\bibfnamefont {N.}~\bibnamefont
  {Brunner}}, \bibinfo {author} {\bibfnamefont {D.}~\bibnamefont {Cavalcanti}},
  \bibinfo {author} {\bibfnamefont {S.}~\bibnamefont {Pironio}}, \bibinfo
  {author} {\bibfnamefont {V.}~\bibnamefont {Scarani}}, \ and\ \bibinfo
  {author} {\bibfnamefont {S.}~\bibnamefont {Wehner}},\ }\href@noop {}
  {\bibfield  {journal} {\bibinfo  {journal} {Rev. Mod. Phys.}\ }\textbf
  {\bibinfo {volume} {86}},\ \bibinfo {pages} {419} (\bibinfo {year}
  {2014})}\BibitemShut {NoStop}%
\bibitem [{\citenamefont {Horodecki}\ \emph {et~al.}(2009)\citenamefont
  {Horodecki}, \citenamefont {Horodecki}, \citenamefont {Horodecki},\ and\
  \citenamefont {Horodecki}}]{horodecki2009quantum}%
  \BibitemOpen
  \bibfield  {author} {\bibinfo {author} {\bibfnamefont {R.}~\bibnamefont
  {Horodecki}}, \bibinfo {author} {\bibfnamefont {P.}~\bibnamefont
  {Horodecki}}, \bibinfo {author} {\bibfnamefont {M.}~\bibnamefont
  {Horodecki}}, \ and\ \bibinfo {author} {\bibfnamefont {K.}~\bibnamefont
  {Horodecki}},\ }\href@noop {} {\bibfield  {journal} {\bibinfo  {journal}
  {Rev. Mod. Phys.}\ }\textbf {\bibinfo {volume} {81}},\ \bibinfo {pages} {865}
  (\bibinfo {year} {2009})}\BibitemShut {NoStop}%
\bibitem [{\citenamefont {Vedral}(2014)}]{vedralReview}%
  \BibitemOpen
  \bibfield  {author} {\bibinfo {author} {\bibfnamefont {V.}~\bibnamefont
  {Vedral}},\ }\href@noop {} {\bibfield  {journal} {\bibinfo  {journal} {Nat.
  Phys.}\ }\textbf {\bibinfo {volume} {10}},\ \bibinfo {pages} {256} (\bibinfo
  {year} {2014})}\BibitemShut {NoStop}%
\bibitem [{\citenamefont {Walter}\ \emph {et~al.}()\citenamefont {Walter},
  \citenamefont {Gross},\ and\ \citenamefont {Eisert}}]{eisertReview}%
  \BibitemOpen
  \bibfield  {author} {\bibinfo {author} {\bibfnamefont {M.}~\bibnamefont
  {Walter}}, \bibinfo {author} {\bibfnamefont {D.}~\bibnamefont {Gross}}, \
  and\ \bibinfo {author} {\bibfnamefont {J.}~\bibnamefont {Eisert}},\
  }\href@noop {} {\bibinfo  {journal} {arXiv:1612.02437 [quant-ph]}\
  }\BibitemShut {NoStop}%
\bibitem [{\citenamefont {Bengtsson}\ and\ \citenamefont
  {Zyczkowski}()}]{bengtssonChapter}%
  \BibitemOpen
\bibfield  {journal} {  }\bibfield  {author} {\bibinfo {author} {\bibfnamefont
  {I.}~\bibnamefont {Bengtsson}}\ and\ \bibinfo {author} {\bibfnamefont
  {K.}~\bibnamefont {Zyczkowski}},\ }\href@noop {} {\bibinfo  {journal}
  {arXiv:1612.07747 [quant-ph]}\ }\BibitemShut {NoStop}%
\bibitem [{\citenamefont {Briegel}\ and\ \citenamefont
  {Raussendorf}(2001)}]{PhysRevLett.86.910}%
  \BibitemOpen
\bibfield  {journal} {  }\bibfield  {author} {\bibinfo {author} {\bibfnamefont
  {H.~J.}\ \bibnamefont {Briegel}}\ and\ \bibinfo {author} {\bibfnamefont
  {R.}~\bibnamefont {Raussendorf}},\ }\href@noop {} {\bibfield  {journal}
  {\bibinfo  {journal} {Phys. Rev. Lett.}\ }\textbf {\bibinfo {volume} {86}},\
  \bibinfo {pages} {910} (\bibinfo {year} {2001})}\BibitemShut {NoStop}%
\bibitem [{\citenamefont {Dicke}(1954)}]{Dicke1954}%
  \BibitemOpen
  \bibfield  {author} {\bibinfo {author} {\bibfnamefont {R.~H.}\ \bibnamefont
  {Dicke}},\ }\href@noop {} {\bibfield  {journal} {\bibinfo  {journal} {Phys.
  Rev.}\ }\textbf {\bibinfo {volume} {93}},\ \bibinfo {pages} {99} (\bibinfo
  {year} {1954})}\BibitemShut {NoStop}%
\bibitem [{\citenamefont {D\"ur}\ \emph {et~al.}(2000)\citenamefont {D\"ur},
  \citenamefont {Vidal},\ and\ \citenamefont {Cirac}}]{ciracPRA2000}%
  \BibitemOpen
  \bibfield  {author} {\bibinfo {author} {\bibfnamefont {W.}~\bibnamefont
  {D\"ur}}, \bibinfo {author} {\bibfnamefont {G.}~\bibnamefont {Vidal}}, \ and\
  \bibinfo {author} {\bibfnamefont {J.~I.}\ \bibnamefont {Cirac}},\ }\href@noop
  {} {\bibfield  {journal} {\bibinfo  {journal} {Phys. Rev. A}\ }\textbf
  {\bibinfo {volume} {62}},\ \bibinfo {pages} {062314} (\bibinfo {year}
  {2000})}\BibitemShut {NoStop}%
\bibitem [{\citenamefont {Tashima}\ \emph {et~al.}(2016)\citenamefont
  {Tashima}, \citenamefont {Tame}, \citenamefont {{\"{O}}zdemir}, \citenamefont
  {Nori}, \citenamefont {Koashi},\ and\ \citenamefont
  {Weinfurter}}]{noriPRA2016}%
  \BibitemOpen
  \bibfield  {author} {\bibinfo {author} {\bibfnamefont {T.}~\bibnamefont
  {Tashima}}, \bibinfo {author} {\bibfnamefont {M.~S.}\ \bibnamefont {Tame}},
  \bibinfo {author} {\bibfnamefont {S.~K.}\ \bibnamefont {{\"{O}}zdemir}},
  \bibinfo {author} {\bibfnamefont {F.}~\bibnamefont {Nori}}, \bibinfo {author}
  {\bibfnamefont {M.}~\bibnamefont {Koashi}}, \ and\ \bibinfo {author}
  {\bibfnamefont {H.}~\bibnamefont {Weinfurter}},\ }\href@noop {} {\bibfield
  {journal} {\bibinfo  {journal} {Phys. Rev. A}\ }\textbf {\bibinfo {volume}
  {94}},\ \bibinfo {pages} {052309} (\bibinfo {year} {2016})}\BibitemShut
  {NoStop}%
\bibitem [{\citenamefont {Regula}\ and\ \citenamefont
  {Adesso}(2016)}]{PhysRevLett.116.070504}%
  \BibitemOpen
  \bibfield  {author} {\bibinfo {author} {\bibfnamefont {B.}~\bibnamefont
  {Regula}}\ and\ \bibinfo {author} {\bibfnamefont {G.}~\bibnamefont
  {Adesso}},\ }\href@noop {} {\bibfield  {journal} {\bibinfo  {journal} {Phys.
  Rev. Lett.}\ }\textbf {\bibinfo {volume} {116}},\ \bibinfo {pages} {070504}
  (\bibinfo {year} {2016})}\BibitemShut {NoStop}%
\bibitem [{\citenamefont {D{\"{u}}r}(2001)}]{durPRA2001}%
  \BibitemOpen
  \bibfield  {author} {\bibinfo {author} {\bibfnamefont {W.}~\bibnamefont
  {D{\"{u}}r}},\ }\href@noop {} {\bibfield  {journal} {\bibinfo  {journal}
  {Phys. Rev. A}\ }\textbf {\bibinfo {volume} {63}},\ \bibinfo {pages} {020303}
  (\bibinfo {year} {2001})}\BibitemShut {NoStop}%
\bibitem [{\citenamefont {Sen(De)}\ \emph {et~al.}(2003)\citenamefont
  {Sen(De)}, \citenamefont {Sen}, \citenamefont {Wie\ifmmode~\acute{s}\else
  \'{s}\fi{}niak}, \citenamefont {Kaszlikowski},\ and\ \citenamefont
  {\ifmmode~\dot{Z}\else \.{Z}\fi{}ukowski}}]{PhysRevA.68.062306}%
  \BibitemOpen
  \bibfield  {author} {\bibinfo {author} {\bibfnamefont {A.}~\bibnamefont
  {Sen(De)}}, \bibinfo {author} {\bibfnamefont {U.}~\bibnamefont {Sen}},
  \bibinfo {author} {\bibfnamefont {M.}~\bibnamefont
  {Wie\ifmmode~\acute{s}\else \'{s}\fi{}niak}}, \bibinfo {author}
  {\bibfnamefont {D.}~\bibnamefont {Kaszlikowski}}, \ and\ \bibinfo {author}
  {\bibfnamefont {M.}~\bibnamefont {\ifmmode~\dot{Z}\else \.{Z}\fi{}ukowski}},\
  }\href@noop {} {\bibfield  {journal} {\bibinfo  {journal} {Phys. Rev. A}\
  }\textbf {\bibinfo {volume} {68}},\ \bibinfo {pages} {062306} (\bibinfo
  {year} {2003})}\BibitemShut {NoStop}%
\bibitem [{\citenamefont {{D'}Hondt}\ and\ \citenamefont
  {Panangaden}(2006)}]{computeW2006}%
  \BibitemOpen
  \bibfield  {author} {\bibinfo {author} {\bibfnamefont {E.}~\bibnamefont
  {{D'}Hondt}}\ and\ \bibinfo {author} {\bibfnamefont {P.}~\bibnamefont
  {Panangaden}},\ }\href@noop {} {\bibfield  {journal} {\bibinfo  {journal}
  {Quantum Inf. Comput.}\ }\textbf {\bibinfo {volume} {6}},\ \bibinfo {pages}
  {173} (\bibinfo {year} {2006})}\BibitemShut {NoStop}%
\bibitem [{\citenamefont {Buhrman}\ \emph {et~al.}(1999)\citenamefont
  {Buhrman}, \citenamefont {van Dam}, \citenamefont {H\o{}yer},\ and\
  \citenamefont {Tapp}}]{PhysRevA.60.2737}%
  \BibitemOpen
  \bibfield  {author} {\bibinfo {author} {\bibfnamefont {H.}~\bibnamefont
  {Buhrman}}, \bibinfo {author} {\bibfnamefont {W.}~\bibnamefont {van Dam}},
  \bibinfo {author} {\bibfnamefont {P.}~\bibnamefont {H\o{}yer}}, \ and\
  \bibinfo {author} {\bibfnamefont {A.}~\bibnamefont {Tapp}},\ }\href@noop {}
  {\bibfield  {journal} {\bibinfo  {journal} {Phys. Rev. A}\ }\textbf {\bibinfo
  {volume} {60}},\ \bibinfo {pages} {2737} (\bibinfo {year}
  {1999})}\BibitemShut {NoStop}%
\bibitem [{\citenamefont {Jian}\ \emph {et~al.}(2007)\citenamefont {Jian},
  \citenamefont {Quan},\ and\ \citenamefont {Chao-Jing}}]{wangWsecure}%
  \BibitemOpen
  \bibfield  {author} {\bibinfo {author} {\bibfnamefont {W.}~\bibnamefont
  {Jian}}, \bibinfo {author} {\bibfnamefont {Z.}~\bibnamefont {Quan}}, \ and\
  \bibinfo {author} {\bibfnamefont {T.}~\bibnamefont {Chao-Jing}},\ }\href@noop
  {} {\bibfield  {journal} {\bibinfo  {journal} {Commun. Theor. Phys.}\
  }\textbf {\bibinfo {volume} {48}},\ \bibinfo {pages} {637} (\bibinfo {year}
  {2007})}\BibitemShut {NoStop}%
\bibitem [{\citenamefont {Joo}\ \emph {et~al.}(2005)\citenamefont {Joo},
  \citenamefont {Y.-J.Park}, \citenamefont {Lee}, \citenamefont {Jang},\ and\
  \citenamefont {Kim}}]{jooWsecure}%
  \BibitemOpen
  \bibfield  {author} {\bibinfo {author} {\bibfnamefont {J.}~\bibnamefont
  {Joo}}, \bibinfo {author} {\bibnamefont {Y.-J.Park}}, \bibinfo {author}
  {\bibfnamefont {J.}~\bibnamefont {Lee}}, \bibinfo {author} {\bibfnamefont
  {J.}~\bibnamefont {Jang}}, \ and\ \bibinfo {author} {\bibfnamefont
  {J.}~\bibnamefont {Kim}},\ }\href@noop {} {\bibfield  {journal} {\bibinfo
  {journal} {J. Korean Phys. Soc.}\ }\textbf {\bibinfo {volume} {46}},\
  \bibinfo {pages} {763} (\bibinfo {year} {2005})}\BibitemShut {NoStop}%
\bibitem [{\citenamefont {Li}\ \emph {et~al.}(2008)\citenamefont {Li},
  \citenamefont {Xiao-Ming}, \citenamefont {Ya-Jun},\ and\ \citenamefont
  {Feng}}]{dongWsecure}%
  \BibitemOpen
  \bibfield  {author} {\bibinfo {author} {\bibfnamefont {D.}~\bibnamefont
  {Li}}, \bibinfo {author} {\bibfnamefont {X.}~\bibnamefont {Xiao-Ming}},
  \bibinfo {author} {\bibfnamefont {G.}~\bibnamefont {Ya-Jun}}, \ and\ \bibinfo
  {author} {\bibfnamefont {C.}~\bibnamefont {Feng}},\ }\href@noop {} {\bibfield
   {journal} {\bibinfo  {journal} {Commun. Theor. Phys.}\ }\textbf {\bibinfo
  {volume} {50}},\ \bibinfo {pages} {359} (\bibinfo {year} {2008})}\BibitemShut
  {NoStop}%
\bibitem [{\citenamefont {Cao}\ and\ \citenamefont {Song}(2006)}]{caoWsecure}%
  \BibitemOpen
  \bibfield  {author} {\bibinfo {author} {\bibfnamefont {H.-J.}\ \bibnamefont
  {Cao}}\ and\ \bibinfo {author} {\bibfnamefont {H.-S.}\ \bibnamefont {Song}},\
  }\href@noop {} {\bibfield  {journal} {\bibinfo  {journal} {Phys. Scr.}\
  }\textbf {\bibinfo {volume} {74}},\ \bibinfo {pages} {572} (\bibinfo {year}
  {2006})}\BibitemShut {NoStop}%
\bibitem [{\citenamefont {Gorbachev}\ \emph {et~al.}(2003)\citenamefont
  {Gorbachev}, \citenamefont {Trubilko}, \citenamefont {Rodichkina},\ and\
  \citenamefont {Zhiliba}}]{Gorbachev2003267}%
  \BibitemOpen
  \bibfield  {author} {\bibinfo {author} {\bibfnamefont {V.~N.}\ \bibnamefont
  {Gorbachev}}, \bibinfo {author} {\bibfnamefont {A.~I.}\ \bibnamefont
  {Trubilko}}, \bibinfo {author} {\bibfnamefont {A.~A.}\ \bibnamefont
  {Rodichkina}}, \ and\ \bibinfo {author} {\bibfnamefont {A.~I.}\ \bibnamefont
  {Zhiliba}},\ }\href@noop {} {\bibfield  {journal} {\bibinfo  {journal} {Phys.
  Lett. A}\ }\textbf {\bibinfo {volume} {314}},\ \bibinfo {pages} {267 }
  (\bibinfo {year} {2003})}\BibitemShut {NoStop}%
\bibitem [{\citenamefont {Joo}\ \emph {et~al.}(2003)\citenamefont {Joo},
  \citenamefont {Park}, \citenamefont {Oh},\ and\ \citenamefont
  {Kim}}]{jooWteleport}%
  \BibitemOpen
  \bibfield  {author} {\bibinfo {author} {\bibfnamefont {J.}~\bibnamefont
  {Joo}}, \bibinfo {author} {\bibfnamefont {Y.-J.}\ \bibnamefont {Park}},
  \bibinfo {author} {\bibfnamefont {S.}~\bibnamefont {Oh}}, \ and\ \bibinfo
  {author} {\bibfnamefont {J.}~\bibnamefont {Kim}},\ }\href@noop {} {\bibfield
  {journal} {\bibinfo  {journal} {New J. Phys.}\ }\textbf {\bibinfo {volume}
  {5}},\ \bibinfo {pages} {136} (\bibinfo {year} {2003})}\BibitemShut {NoStop}%
\bibitem [{\citenamefont {Dag}\ \emph {et~al.}(2016)\citenamefont {Dag},
  \citenamefont {Niedenzu}, \citenamefont {Mustecaplioglu},\ and\ \citenamefont
  {Kurizki}}]{dagEntropy}%
  \BibitemOpen
  \bibfield  {author} {\bibinfo {author} {\bibfnamefont {C.~B.}\ \bibnamefont
  {Dag}}, \bibinfo {author} {\bibfnamefont {W.}~\bibnamefont {Niedenzu}},
  \bibinfo {author} {\bibfnamefont {O.~E.}\ \bibnamefont {Mustecaplioglu}}, \
  and\ \bibinfo {author} {\bibfnamefont {G.}~\bibnamefont {Kurizki}},\
  }\href@noop {} {\bibfield  {journal} {\bibinfo  {journal} {Entropy}\ }\textbf
  {\bibinfo {volume} {18}},\ \bibinfo {pages} {244} (\bibinfo {year}
  {2016})}\BibitemShut {NoStop}%
\bibitem [{\citenamefont {Sharma}\ \emph {et~al.}(2008)\citenamefont {Sharma},
  \citenamefont {Almeida},\ and\ \citenamefont {Sharma}}]{sharmaJPB2008}%
  \BibitemOpen
  \bibfield  {author} {\bibinfo {author} {\bibfnamefont {S.~S.}\ \bibnamefont
  {Sharma}}, \bibinfo {author} {\bibfnamefont {E.}~\bibnamefont {Almeida}}, \
  and\ \bibinfo {author} {\bibfnamefont {N.~K.}\ \bibnamefont {Sharma}},\
  }\href@noop {} {\bibfield  {journal} {\bibinfo  {journal} {J. Phys. B: At.
  Mol. Opt. Phys.}\ }\textbf {\bibinfo {volume} {41}},\ \bibinfo {pages}
  {165503} (\bibinfo {year} {2008})}\BibitemShut {NoStop}%
\bibitem [{\citenamefont {Liu}\ \emph {et~al.}(2014)\citenamefont {Liu},
  \citenamefont {Yu}, \citenamefont {Li},\ and\ \citenamefont
  {Wu}}]{liu2014JAP}%
  \BibitemOpen
  \bibfield  {author} {\bibinfo {author} {\bibfnamefont {S.}~\bibnamefont
  {Liu}}, \bibinfo {author} {\bibfnamefont {R.}~\bibnamefont {Yu}}, \bibinfo
  {author} {\bibfnamefont {J.}~\bibnamefont {Li}}, \ and\ \bibinfo {author}
  {\bibfnamefont {Y.}~\bibnamefont {Wu}},\ }\href@noop {} {\bibfield  {journal}
  {\bibinfo  {journal} {J. App. Phys.}\ }\textbf {\bibinfo {volume} {115}},\
  \bibinfo {pages} {134312} (\bibinfo {year} {2014})}\BibitemShut {NoStop}%
\bibitem [{\citenamefont {Yu}\ \emph {et~al.}(2007)\citenamefont {Yu},
  \citenamefont {Yi}, \citenamefont {Song},\ and\ \citenamefont
  {Mei}}]{yuPRA2007}%
  \BibitemOpen
  \bibfield  {author} {\bibinfo {author} {\bibfnamefont {C.-S.}\ \bibnamefont
  {Yu}}, \bibinfo {author} {\bibfnamefont {X.~X.}\ \bibnamefont {Yi}}, \bibinfo
  {author} {\bibfnamefont {H.-S.}\ \bibnamefont {Song}}, \ and\ \bibinfo
  {author} {\bibfnamefont {D.}~\bibnamefont {Mei}},\ }\href@noop {} {\bibfield
  {journal} {\bibinfo  {journal} {Phys. Rev. A}\ }\textbf {\bibinfo {volume}
  {75}},\ \bibinfo {pages} {044301} (\bibinfo {year} {2007})}\BibitemShut
  {NoStop}%
\bibitem [{\citenamefont {Zou}\ \emph {et~al.}(2002)\citenamefont {Zou},
  \citenamefont {Pahlke},\ and\ \citenamefont {Mathis}}]{zouPRA2002}%
  \BibitemOpen
  \bibfield  {author} {\bibinfo {author} {\bibfnamefont {X.~B.}\ \bibnamefont
  {Zou}}, \bibinfo {author} {\bibfnamefont {K.}~\bibnamefont {Pahlke}}, \ and\
  \bibinfo {author} {\bibfnamefont {W.}~\bibnamefont {Mathis}},\ }\href@noop {}
  {\bibfield  {journal} {\bibinfo  {journal} {Phys. Rev. A}\ }\textbf {\bibinfo
  {volume} {66}},\ \bibinfo {pages} {044302} (\bibinfo {year}
  {2002})}\BibitemShut {NoStop}%
\bibitem [{\citenamefont {Yamamoto}\ \emph {et~al.}(2002)\citenamefont
  {Yamamoto}, \citenamefont {Tamaki}, \citenamefont {Koashi},\ and\
  \citenamefont {Imoto}}]{PhysRevA.66.064301}%
  \BibitemOpen
  \bibfield  {author} {\bibinfo {author} {\bibfnamefont {T.}~\bibnamefont
  {Yamamoto}}, \bibinfo {author} {\bibfnamefont {K.}~\bibnamefont {Tamaki}},
  \bibinfo {author} {\bibfnamefont {M.}~\bibnamefont {Koashi}}, \ and\ \bibinfo
  {author} {\bibfnamefont {N.}~\bibnamefont {Imoto}},\ }\href@noop {}
  {\bibfield  {journal} {\bibinfo  {journal} {Phys. Rev. A}\ }\textbf {\bibinfo
  {volume} {66}},\ \bibinfo {pages} {064301} (\bibinfo {year}
  {2002})}\BibitemShut {NoStop}%
\bibitem [{\citenamefont {Kang}\ \emph {et~al.}(2016)\citenamefont {Kang},
  \citenamefont {Chen}, \citenamefont {amd B.-H.~Huang}, \citenamefont {Song},\
  and\ \citenamefont {Xia}}]{kangSciRep2016}%
  \BibitemOpen
  \bibfield  {author} {\bibinfo {author} {\bibfnamefont {Y.-H.}\ \bibnamefont
  {Kang}}, \bibinfo {author} {\bibfnamefont {Y.-H.}\ \bibnamefont {Chen}},
  \bibinfo {author} {\bibfnamefont {Q.-C.~W.}\ \bibnamefont {amd B.-H.~Huang}},
  \bibinfo {author} {\bibfnamefont {J.}~\bibnamefont {Song}}, \ and\ \bibinfo
  {author} {\bibfnamefont {Y.}~\bibnamefont {Xia}},\ }\href@noop {} {\bibfield
  {journal} {\bibinfo  {journal} {Sci. Rep.}\ }\textbf {\bibinfo {volume}
  {6}},\ \bibinfo {pages} {36737} (\bibinfo {year} {2016})}\BibitemShut
  {NoStop}%
\bibitem [{\citenamefont {Biswas}\ and\ \citenamefont
  {Agarwal}(2004)}]{biswasJMP}%
  \BibitemOpen
  \bibfield  {author} {\bibinfo {author} {\bibfnamefont {A.}~\bibnamefont
  {Biswas}}\ and\ \bibinfo {author} {\bibfnamefont {G.~S.}\ \bibnamefont
  {Agarwal}},\ }\href@noop {} {\bibfield  {journal} {\bibinfo  {journal} {J.
  Mod. Opt.}\ }\textbf {\bibinfo {volume} {51}},\ \bibinfo {pages} {1627}
  (\bibinfo {year} {2004})}\BibitemShut {NoStop}%
\bibitem [{\citenamefont {Sweke}\ \emph {et~al.}(2013)\citenamefont {Sweke},
  \citenamefont {Sinayskiy},\ and\ \citenamefont {Petruccione}}]{swekePRA}%
  \BibitemOpen
  \bibfield  {author} {\bibinfo {author} {\bibfnamefont {R.}~\bibnamefont
  {Sweke}}, \bibinfo {author} {\bibfnamefont {I.}~\bibnamefont {Sinayskiy}}, \
  and\ \bibinfo {author} {\bibfnamefont {F.}~\bibnamefont {Petruccione}},\
  }\href@noop {} {\bibfield  {journal} {\bibinfo  {journal} {Phys. Rev. A}\
  }\textbf {\bibinfo {volume} {87}},\ \bibinfo {pages} {042323} (\bibinfo
  {year} {2013})}\BibitemShut {NoStop}%
\bibitem [{\citenamefont {Gao}\ \emph {et~al.}(2013)\citenamefont {Gao},
  \citenamefont {Zhou}, \citenamefont {Zou}, \citenamefont {Peng},\ and\
  \citenamefont {Du}}]{gaoPRA}%
  \BibitemOpen
  \bibfield  {author} {\bibinfo {author} {\bibfnamefont {Y.}~\bibnamefont
  {Gao}}, \bibinfo {author} {\bibfnamefont {H.}~\bibnamefont {Zhou}}, \bibinfo
  {author} {\bibfnamefont {D.}~\bibnamefont {Zou}}, \bibinfo {author}
  {\bibfnamefont {X.}~\bibnamefont {Peng}}, \ and\ \bibinfo {author}
  {\bibfnamefont {J.}~\bibnamefont {Du}},\ }\href@noop {} {\bibfield  {journal}
  {\bibinfo  {journal} {Phys. Rev. A}\ }\textbf {\bibinfo {volume} {87}},\
  \bibinfo {pages} {032335} (\bibinfo {year} {2013})}\BibitemShut {NoStop}%
\bibitem [{\citenamefont {Moreno}\ \emph {et~al.}(2016)\citenamefont {Moreno},
  \citenamefont {Cunha},\ and\ \citenamefont {Parisio}}]{Moreno2016}%
  \BibitemOpen
  \bibfield  {author} {\bibinfo {author} {\bibfnamefont {M.~G.~M.}\
  \bibnamefont {Moreno}}, \bibinfo {author} {\bibfnamefont {M.~M.}\
  \bibnamefont {Cunha}}, \ and\ \bibinfo {author} {\bibfnamefont
  {F.}~\bibnamefont {Parisio}},\ }\href@noop {} {\bibfield  {journal} {\bibinfo
   {journal} {Quantum Inform. Process.}\ }\textbf {\bibinfo {volume} {15}},\
  \bibinfo {pages} {3869} (\bibinfo {year} {2016})}\BibitemShut {NoStop}%
\bibitem [{\citenamefont {Krenn}\ \emph {et~al.}(2017)\citenamefont {Krenn},
  \citenamefont {Hochrainer}, \citenamefont {Lahiri},\ and\ \citenamefont
  {Zeilinger}}]{krenn2017}%
  \BibitemOpen
  \bibfield  {author} {\bibinfo {author} {\bibfnamefont {M.}~\bibnamefont
  {Krenn}}, \bibinfo {author} {\bibfnamefont {A.}~\bibnamefont {Hochrainer}},
  \bibinfo {author} {\bibfnamefont {M.}~\bibnamefont {Lahiri}}, \ and\ \bibinfo
  {author} {\bibfnamefont {A.}~\bibnamefont {Zeilinger}},\ }\href@noop {}
  {\bibfield  {journal} {\bibinfo  {journal} {Phys. Rev. Lett.}\ }\textbf
  {\bibinfo {volume} {118}},\ \bibinfo {pages} {080401} (\bibinfo {year}
  {2017})}\BibitemShut {NoStop}%
\bibitem [{\citenamefont {Li}(2006)}]{gaoxiangPRA}%
  \BibitemOpen
  \bibfield  {author} {\bibinfo {author} {\bibfnamefont {G.-X.}\ \bibnamefont
  {Li}},\ }\href@noop {} {\bibfield  {journal} {\bibinfo  {journal} {Phys. Rev.
  A}\ }\textbf {\bibinfo {volume} {74}},\ \bibinfo {pages} {055801} (\bibinfo
  {year} {2006})}\BibitemShut {NoStop}%
\bibitem [{\citenamefont {He}\ \emph {et~al.}(2014)\citenamefont {He},
  \citenamefont {Su}, \citenamefont {Zhang},\ and\ \citenamefont
  {Yang}}]{He2014QIP}%
  \BibitemOpen
  \bibfield  {author} {\bibinfo {author} {\bibfnamefont {X.-L.}\ \bibnamefont
  {He}}, \bibinfo {author} {\bibfnamefont {Q.-P.}\ \bibnamefont {Su}}, \bibinfo
  {author} {\bibfnamefont {F.-Y.}\ \bibnamefont {Zhang}}, \ and\ \bibinfo
  {author} {\bibfnamefont {C.-P.}\ \bibnamefont {Yang}},\ }\href@noop {}
  {\bibfield  {journal} {\bibinfo  {journal} {Quantum Inf. Process.}\ }\textbf
  {\bibinfo {volume} {13}},\ \bibinfo {pages} {1381} (\bibinfo {year}
  {2014})}\BibitemShut {NoStop}%
\bibitem [{\citenamefont {Yesilyurt}\ \emph {et~al.}(2016)\citenamefont
  {Yesilyurt}, \citenamefont {Bugu}, \citenamefont {Ozaydin}, \citenamefont
  {Altintas}, \citenamefont {Tame}, \citenamefont {Yang},\ and\ \citenamefont
  {{\"{O}}zdemir}}]{Yesilyurt:16}%
  \BibitemOpen
  \bibfield  {author} {\bibinfo {author} {\bibfnamefont {C.}~\bibnamefont
  {Yesilyurt}}, \bibinfo {author} {\bibfnamefont {S.}~\bibnamefont {Bugu}},
  \bibinfo {author} {\bibfnamefont {F.}~\bibnamefont {Ozaydin}}, \bibinfo
  {author} {\bibfnamefont {A.~A.}\ \bibnamefont {Altintas}}, \bibinfo {author}
  {\bibfnamefont {M.}~\bibnamefont {Tame}}, \bibinfo {author} {\bibfnamefont
  {L.}~\bibnamefont {Yang}}, \ and\ \bibinfo {author} {\bibfnamefont
  {{\c{S}}.~K.}\ \bibnamefont {{\"{O}}zdemir}},\ }\href@noop {} {\bibfield
  {journal} {\bibinfo  {journal} {J. Opt. Soc. Am. B}\ }\textbf {\bibinfo
  {volume} {33}},\ \bibinfo {pages} {2313} (\bibinfo {year}
  {2016})}\BibitemShut {NoStop}%
\bibitem [{\citenamefont {Xue}\ and\ \citenamefont
  {Guo}(2003)}]{PhysRevA.67.034302}%
  \BibitemOpen
  \bibfield  {author} {\bibinfo {author} {\bibfnamefont {P.}~\bibnamefont
  {Xue}}\ and\ \bibinfo {author} {\bibfnamefont {G.-C.}\ \bibnamefont {Guo}},\
  }\href@noop {} {\bibfield  {journal} {\bibinfo  {journal} {Phys. Rev. A}\
  }\textbf {\bibinfo {volume} {67}},\ \bibinfo {pages} {034302} (\bibinfo
  {year} {2003})}\BibitemShut {NoStop}%
\bibitem [{\citenamefont {Wang}\ \emph {et~al.}(2003)\citenamefont {Wang},
  \citenamefont {Feng},\ and\ \citenamefont {Sanders}}]{PhysRevA.67.022302}%
  \BibitemOpen
  \bibfield  {author} {\bibinfo {author} {\bibfnamefont {X.}~\bibnamefont
  {Wang}}, \bibinfo {author} {\bibfnamefont {M.}~\bibnamefont {Feng}}, \ and\
  \bibinfo {author} {\bibfnamefont {B.~C.}\ \bibnamefont {Sanders}},\
  }\href@noop {} {\bibfield  {journal} {\bibinfo  {journal} {Phys. Rev. A}\
  }\textbf {\bibinfo {volume} {67}},\ \bibinfo {pages} {022302} (\bibinfo
  {year} {2003})}\BibitemShut {NoStop}%
\bibitem [{\citenamefont {Song}\ \emph {et~al.}(2005)\citenamefont {Song},
  \citenamefont {Zhou},\ and\ \citenamefont {Guo}}]{PhysRevA.71.052310}%
  \BibitemOpen
  \bibfield  {author} {\bibinfo {author} {\bibfnamefont {K.-H.}\ \bibnamefont
  {Song}}, \bibinfo {author} {\bibfnamefont {Z.-W.}\ \bibnamefont {Zhou}}, \
  and\ \bibinfo {author} {\bibfnamefont {G.-C.}\ \bibnamefont {Guo}},\
  }\href@noop {} {\bibfield  {journal} {\bibinfo  {journal} {Phys. Rev. A}\
  }\textbf {\bibinfo {volume} {71}},\ \bibinfo {pages} {052310} (\bibinfo
  {year} {2005})}\BibitemShut {NoStop}%
\bibitem [{\citenamefont {Song}\ \emph {et~al.}(2007)\citenamefont {Song},
  \citenamefont {Xiang}, \citenamefont {Liu},\ and\ \citenamefont
  {Lu}}]{songPRA2007}%
  \BibitemOpen
  \bibfield  {author} {\bibinfo {author} {\bibfnamefont {K.~H.}\ \bibnamefont
  {Song}}, \bibinfo {author} {\bibfnamefont {S.~H.}\ \bibnamefont {Xiang}},
  \bibinfo {author} {\bibfnamefont {Q.}~\bibnamefont {Liu}}, \ and\ \bibinfo
  {author} {\bibfnamefont {D.~H.}\ \bibnamefont {Lu}},\ }\href@noop {}
  {\bibfield  {journal} {\bibinfo  {journal} {Phys. Rev. A}\ }\textbf {\bibinfo
  {volume} {75}},\ \bibinfo {pages} {032347} (\bibinfo {year}
  {2007})}\BibitemShut {NoStop}%
\bibitem [{\citenamefont {Zhang}\ \emph {et~al.}(2006)\citenamefont {Zhang},
  \citenamefont {Gao},\ and\ \citenamefont {Feng}}]{zhangPRA2006}%
  \BibitemOpen
  \bibfield  {author} {\bibinfo {author} {\bibfnamefont {X.~L.}\ \bibnamefont
  {Zhang}}, \bibinfo {author} {\bibfnamefont {K.~L.}\ \bibnamefont {Gao}}, \
  and\ \bibinfo {author} {\bibfnamefont {M.}~\bibnamefont {Feng}},\ }\href@noop
  {} {\bibfield  {journal} {\bibinfo  {journal} {Phys. Rev. A}\ }\textbf
  {\bibinfo {volume} {74}},\ \bibinfo {pages} {024303} (\bibinfo {year}
  {2006})}\BibitemShut {NoStop}%
\bibitem [{\citenamefont {Deng}\ \emph {et~al.}(2006)\citenamefont {Deng},
  \citenamefont {Gao},\ and\ \citenamefont {Feng}}]{dengPRA2006}%
  \BibitemOpen
  \bibfield  {author} {\bibinfo {author} {\bibfnamefont {Z.~J.}\ \bibnamefont
  {Deng}}, \bibinfo {author} {\bibfnamefont {K.~L.}\ \bibnamefont {Gao}}, \
  and\ \bibinfo {author} {\bibfnamefont {M.}~\bibnamefont {Feng}},\ }\href@noop
  {} {\bibfield  {journal} {\bibinfo  {journal} {Phys. Rev. A}\ }\textbf
  {\bibinfo {volume} {74}},\ \bibinfo {pages} {064303} (\bibinfo {year}
  {2006})}\BibitemShut {NoStop}%
\bibitem [{\citenamefont {Ikuta}\ \emph {et~al.}(2011)\citenamefont {Ikuta},
  \citenamefont {Tashima}, \citenamefont {Yamamoto}, \citenamefont {Koashi},\
  and\ \citenamefont {Imoto}}]{ikuta2011}%
  \BibitemOpen
  \bibfield  {author} {\bibinfo {author} {\bibfnamefont {R.}~\bibnamefont
  {Ikuta}}, \bibinfo {author} {\bibfnamefont {T.}~\bibnamefont {Tashima}},
  \bibinfo {author} {\bibfnamefont {T.}~\bibnamefont {Yamamoto}}, \bibinfo
  {author} {\bibfnamefont {M.}~\bibnamefont {Koashi}}, \ and\ \bibinfo {author}
  {\bibfnamefont {N.}~\bibnamefont {Imoto}},\ }\href@noop {} {\bibfield
  {journal} {\bibinfo  {journal} {Phys. Rev. A}\ }\textbf {\bibinfo {volume}
  {83}},\ \bibinfo {pages} {012314} (\bibinfo {year} {2011})}\BibitemShut
  {NoStop}%
\bibitem [{\citenamefont {{\"{O}}zdemir}\ \emph {et~al.}(2011)\citenamefont
  {{\"{O}}zdemir}, \citenamefont {Matsunaga}, \citenamefont {Tashima},
  \citenamefont {Yamamoto}, \citenamefont {Koashi},\ and\ \citenamefont
  {Imoto}}]{imotoNJP}%
  \BibitemOpen
  \bibfield  {author} {\bibinfo {author} {\bibfnamefont {S.~K.}\ \bibnamefont
  {{\"{O}}zdemir}}, \bibinfo {author} {\bibfnamefont {E.}~\bibnamefont
  {Matsunaga}}, \bibinfo {author} {\bibfnamefont {T.}~\bibnamefont {Tashima}},
  \bibinfo {author} {\bibfnamefont {T.}~\bibnamefont {Yamamoto}}, \bibinfo
  {author} {\bibfnamefont {M.}~\bibnamefont {Koashi}}, \ and\ \bibinfo {author}
  {\bibfnamefont {N.}~\bibnamefont {Imoto}},\ }\href@noop {} {\bibfield
  {journal} {\bibinfo  {journal} {New J. Phys.}\ }\textbf {\bibinfo {volume}
  {13}},\ \bibinfo {pages} {103003} (\bibinfo {year} {2011})}\BibitemShut
  {NoStop}%
\bibitem [{\citenamefont {Perez-Leija}\ \emph {et~al.}(2013)\citenamefont
  {Perez-Leija}, \citenamefont {Hernandez-Herrejon}, \citenamefont
  {Moya-Cessa}, \citenamefont {Szameit},\ and\ \citenamefont
  {Christodoulides}}]{perezPRA}%
  \BibitemOpen
  \bibfield  {author} {\bibinfo {author} {\bibfnamefont {A.}~\bibnamefont
  {Perez-Leija}}, \bibinfo {author} {\bibfnamefont {J.~C.}\ \bibnamefont
  {Hernandez-Herrejon}}, \bibinfo {author} {\bibfnamefont {H.}~\bibnamefont
  {Moya-Cessa}}, \bibinfo {author} {\bibfnamefont {A.}~\bibnamefont {Szameit}},
  \ and\ \bibinfo {author} {\bibfnamefont {D.~N.}\ \bibnamefont
  {Christodoulides}},\ }\href@noop {} {\bibfield  {journal} {\bibinfo
  {journal} {Phys. Rev. A}\ }\textbf {\bibinfo {volume} {87}},\ \bibinfo
  {pages} {013842} (\bibinfo {year} {2013})}\BibitemShut {NoStop}%
\bibitem [{\citenamefont {Han}\ \emph {et~al.}(2015)\citenamefont {Han},
  \citenamefont {Hu}, \citenamefont {Guo}, \citenamefont {Wang}, \citenamefont
  {Zhu},\ and\ \citenamefont {Zhang}}]{han2015SciRep}%
  \BibitemOpen
  \bibfield  {author} {\bibinfo {author} {\bibfnamefont {X.}~\bibnamefont
  {Han}}, \bibinfo {author} {\bibfnamefont {S.}~\bibnamefont {Hu}}, \bibinfo
  {author} {\bibfnamefont {Q.}~\bibnamefont {Guo}}, \bibinfo {author}
  {\bibfnamefont {H.-F.}\ \bibnamefont {Wang}}, \bibinfo {author}
  {\bibfnamefont {A.-D.}\ \bibnamefont {Zhu}}, \ and\ \bibinfo {author}
  {\bibfnamefont {S.}~\bibnamefont {Zhang}},\ }\href@noop {} {\bibfield
  {journal} {\bibinfo  {journal} {Sci. Rep.}\ }\textbf {\bibinfo {volume}
  {5}},\ \bibinfo {pages} {12790} (\bibinfo {year} {2015})}\BibitemShut
  {NoStop}%
\bibitem [{\citenamefont {Zang}\ \emph {et~al.}(2015)\citenamefont {Zang},
  \citenamefont {Yang}, \citenamefont {Ozaydin}, \citenamefont {Song},\ and\
  \citenamefont {Cao}}]{zang2015SciRep}%
  \BibitemOpen
  \bibfield  {author} {\bibinfo {author} {\bibfnamefont {X.-P.}\ \bibnamefont
  {Zang}}, \bibinfo {author} {\bibfnamefont {M.}~\bibnamefont {Yang}}, \bibinfo
  {author} {\bibfnamefont {F.}~\bibnamefont {Ozaydin}}, \bibinfo {author}
  {\bibfnamefont {W.}~\bibnamefont {Song}}, \ and\ \bibinfo {author}
  {\bibfnamefont {Z.-L.}\ \bibnamefont {Cao}},\ }\href@noop {} {\bibfield
  {journal} {\bibinfo  {journal} {Sci. Rep.}\ }\textbf {\bibinfo {volume}
  {5}},\ \bibinfo {pages} {16245} (\bibinfo {year} {2015})}\BibitemShut
  {NoStop}%
\bibitem [{\citenamefont {Zang}\ \emph {et~al.}(2016)\citenamefont {Zang},
  \citenamefont {Yang}, \citenamefont {Ozaydin}, \citenamefont {Song},\ and\
  \citenamefont {Cao}}]{Zang:16}%
  \BibitemOpen
  \bibfield  {author} {\bibinfo {author} {\bibfnamefont {X.-P.}\ \bibnamefont
  {Zang}}, \bibinfo {author} {\bibfnamefont {M.}~\bibnamefont {Yang}}, \bibinfo
  {author} {\bibfnamefont {F.}~\bibnamefont {Ozaydin}}, \bibinfo {author}
  {\bibfnamefont {W.}~\bibnamefont {Song}}, \ and\ \bibinfo {author}
  {\bibfnamefont {Z.-L.}\ \bibnamefont {Cao}},\ }\href@noop {} {\bibfield
  {journal} {\bibinfo  {journal} {Opt. Express}\ }\textbf {\bibinfo {volume}
  {24}},\ \bibinfo {pages} {12293} (\bibinfo {year} {2016})}\BibitemShut
  {NoStop}%
\bibitem [{\citenamefont {Gr{\"{a}}fe}\ \emph {et~al.}(2014)\citenamefont
  {Gr{\"{a}}fe} \emph {et~al.}}]{grafeNatPhoton}%
  \BibitemOpen
  \bibfield  {author} {\bibinfo {author} {\bibfnamefont {M.}~\bibnamefont
  {Gr{\"{a}}fe}} \emph {et~al.},\ }\href@noop {} {\bibfield  {journal}
  {\bibinfo  {journal} {Nat. Photon.}\ }\textbf {\bibinfo {volume} {8}},\
  \bibinfo {pages} {791} (\bibinfo {year} {2014})}\BibitemShut {NoStop}%
\bibitem [{\citenamefont {Tashima}\ \emph {et~al.}(2009)\citenamefont
  {Tashima}, \citenamefont {Wakatsuki}, \citenamefont {\"Ozdemir},
  \citenamefont {Yamamoto}, \citenamefont {Koashi},\ and\ \citenamefont
  {Imoto}}]{tashima2009}%
  \BibitemOpen
  \bibfield  {author} {\bibinfo {author} {\bibfnamefont {T.}~\bibnamefont
  {Tashima}}, \bibinfo {author} {\bibfnamefont {T.}~\bibnamefont {Wakatsuki}},
  \bibinfo {author} {\bibfnamefont {S.~K.}\ \bibnamefont {\"Ozdemir}}, \bibinfo
  {author} {\bibfnamefont {T.}~\bibnamefont {Yamamoto}}, \bibinfo {author}
  {\bibfnamefont {M.}~\bibnamefont {Koashi}}, \ and\ \bibinfo {author}
  {\bibfnamefont {N.}~\bibnamefont {Imoto}},\ }\href@noop {} {\bibfield
  {journal} {\bibinfo  {journal} {Phys. Rev. Lett.}\ }\textbf {\bibinfo
  {volume} {102}},\ \bibinfo {pages} {130502} (\bibinfo {year}
  {2009})}\BibitemShut {NoStop}%
\bibitem [{\citenamefont {Tashima}\ \emph {et~al.}(2010)\citenamefont
  {Tashima}, \citenamefont {Kitano}, \citenamefont {{\"{O}}zdemir},
  \citenamefont {Yamamoto}, \citenamefont {Koashi},\ and\ \citenamefont
  {Imoto}}]{tashimaPRL}%
  \BibitemOpen
  \bibfield  {author} {\bibinfo {author} {\bibfnamefont {T.}~\bibnamefont
  {Tashima}}, \bibinfo {author} {\bibfnamefont {T.}~\bibnamefont {Kitano}},
  \bibinfo {author} {\bibfnamefont {{\c{S}}.~K.}\ \bibnamefont
  {{\"{O}}zdemir}}, \bibinfo {author} {\bibfnamefont {T.}~\bibnamefont
  {Yamamoto}}, \bibinfo {author} {\bibfnamefont {M.}~\bibnamefont {Koashi}}, \
  and\ \bibinfo {author} {\bibfnamefont {N.}~\bibnamefont {Imoto}},\
  }\href@noop {} {\bibfield  {journal} {\bibinfo  {journal} {Phys. Rev. Lett.}\
  }\textbf {\bibinfo {volume} {105}},\ \bibinfo {pages} {210503} (\bibinfo
  {year} {2010})}\BibitemShut {NoStop}%
\bibitem [{\citenamefont {H{\"{a}}ffner}\ \emph {et~al.}(2005)\citenamefont
  {H{\"{a}}ffner} \emph {et~al.}}]{blattNature2005}%
  \BibitemOpen
  \bibfield  {author} {\bibinfo {author} {\bibfnamefont {H.}~\bibnamefont
  {H{\"{a}}ffner}} \emph {et~al.},\ }\href@noop {} {\bibfield  {journal}
  {\bibinfo  {journal} {Nature}\ }\textbf {\bibinfo {volume} {438}},\ \bibinfo
  {pages} {643} (\bibinfo {year} {2005})}\BibitemShut {NoStop}%
\bibitem [{\citenamefont {Papp}\ \emph {et~al.}(2009)\citenamefont {Papp},
  \citenamefont {Choi}, \citenamefont {Deng}, \citenamefont {Lougovski},
  \citenamefont {{van Enk}},\ and\ \citenamefont {Kimble}}]{4opticalmodes2009}%
  \BibitemOpen
  \bibfield  {author} {\bibinfo {author} {\bibfnamefont {S.~B.}\ \bibnamefont
  {Papp}}, \bibinfo {author} {\bibfnamefont {K.~S.}\ \bibnamefont {Choi}},
  \bibinfo {author} {\bibfnamefont {H.}~\bibnamefont {Deng}}, \bibinfo {author}
  {\bibfnamefont {P.}~\bibnamefont {Lougovski}}, \bibinfo {author}
  {\bibfnamefont {S.}~\bibnamefont {{van Enk}}}, \ and\ \bibinfo {author}
  {\bibfnamefont {H.~J.}\ \bibnamefont {Kimble}},\ }\href@noop {} {\bibfield
  {journal} {\bibinfo  {journal} {Science}\ }\textbf {\bibinfo {volume}
  {324}},\ \bibinfo {pages} {764} (\bibinfo {year} {2009})}\BibitemShut
  {NoStop}%
\bibitem [{\citenamefont {Neeley}\ \emph {et~al.}(2010)\citenamefont {Neeley}
  \emph {et~al.}}]{neeleyNature2010}%
  \BibitemOpen
  \bibfield  {author} {\bibinfo {author} {\bibfnamefont {M.}~\bibnamefont
  {Neeley}} \emph {et~al.},\ }\href@noop {} {\bibfield  {journal} {\bibinfo
  {journal} {Nature}\ }\textbf {\bibinfo {volume} {467}},\ \bibinfo {pages}
  {570} (\bibinfo {year} {2010})}\BibitemShut {NoStop}%
\bibitem [{\citenamefont {Choi}\ \emph {et~al.}(2010)\citenamefont {Choi},
  \citenamefont {Goban}, \citenamefont {Papp}, \citenamefont {van Enk},\ and\
  \citenamefont {Kimble}}]{choiNature2010}%
  \BibitemOpen
  \bibfield  {author} {\bibinfo {author} {\bibfnamefont {K.~S.}\ \bibnamefont
  {Choi}}, \bibinfo {author} {\bibfnamefont {A.}~\bibnamefont {Goban}},
  \bibinfo {author} {\bibfnamefont {S.~B.}\ \bibnamefont {Papp}}, \bibinfo
  {author} {\bibfnamefont {S.~J.}\ \bibnamefont {van Enk}}, \ and\ \bibinfo
  {author} {\bibfnamefont {H.~J.}\ \bibnamefont {Kimble}},\ }\href@noop {}
  {\bibfield  {journal} {\bibinfo  {journal} {Nature}\ }\textbf {\bibinfo
  {volume} {468}},\ \bibinfo {pages} {412} (\bibinfo {year}
  {2010})}\BibitemShut {NoStop}%
\bibitem [{\citenamefont {Altomare}\ \emph {et~al.}(2010)\citenamefont
  {Altomare}, \citenamefont {Park}, \citenamefont {Cicak}, \citenamefont
  {Sillanp{\"{a}\"{a}}}, \citenamefont {Allman}, \citenamefont {Li},
  \citenamefont {Sirois}, \citenamefont {Strong}, \citenamefont {Whittaker},\
  and\ \citenamefont {Simmonds}}]{altomareNatPhys2010}%
  \BibitemOpen
  \bibfield  {author} {\bibinfo {author} {\bibfnamefont {F.}~\bibnamefont
  {Altomare}}, \bibinfo {author} {\bibfnamefont {J.}~\bibnamefont {Park}},
  \bibinfo {author} {\bibfnamefont {K.}~\bibnamefont {Cicak}}, \bibinfo
  {author} {\bibfnamefont {M.}~\bibnamefont {Sillanp{\"{a}\"{a}}}}, \bibinfo
  {author} {\bibfnamefont {M.}~\bibnamefont {Allman}}, \bibinfo {author}
  {\bibfnamefont {D.}~\bibnamefont {Li}}, \bibinfo {author} {\bibfnamefont
  {A.}~\bibnamefont {Sirois}}, \bibinfo {author} {\bibfnamefont
  {J.}~\bibnamefont {Strong}}, \bibinfo {author} {\bibfnamefont
  {J.}~\bibnamefont {Whittaker}}, \ and\ \bibinfo {author} {\bibfnamefont
  {R.}~\bibnamefont {Simmonds}},\ }\href@noop {} {\bibfield  {journal}
  {\bibinfo  {journal} {Nat. Phys.}\ }\textbf {\bibinfo {volume} {6}},\
  \bibinfo {pages} {777} (\bibinfo {year} {2010})}\BibitemShut {NoStop}%
\bibitem [{\citenamefont {Lo~Franco}\ and\ \citenamefont
  {Compagno}(2016)}]{lofranco2015quantum}%
  \BibitemOpen
  \bibfield  {author} {\bibinfo {author} {\bibfnamefont {R.}~\bibnamefont
  {Lo~Franco}}\ and\ \bibinfo {author} {\bibfnamefont {G.}~\bibnamefont
  {Compagno}},\ }\href@noop {} {\bibfield  {journal} {\bibinfo  {journal} {Sci.
  Rep.}\ }\textbf {\bibinfo {volume} {6}},\ \bibinfo {pages} {20603} (\bibinfo
  {year} {2016})}\BibitemShut {NoStop}%
\bibitem [{\citenamefont {Sciara}\ \emph {et~al.}(2017)\citenamefont {Sciara},
  \citenamefont {Lo~Franco},\ and\ \citenamefont {Compagno}}]{sciaraSchmidt}%
  \BibitemOpen
  \bibfield  {author} {\bibinfo {author} {\bibfnamefont {S.}~\bibnamefont
  {Sciara}}, \bibinfo {author} {\bibfnamefont {R.}~\bibnamefont {Lo~Franco}}, \
  and\ \bibinfo {author} {\bibfnamefont {G.}~\bibnamefont {Compagno}},\
  }\href@noop {} {\bibfield  {journal} {\bibinfo  {journal} {Sci. Rep.}\
  }\textbf {\bibinfo {volume} {7}},\ \bibinfo {pages} {44675} (\bibinfo {year}
  {2017})}\BibitemShut {NoStop}%
\bibitem [{\citenamefont {Tichy}\ \emph {et~al.}(2013)\citenamefont {Tichy},
  \citenamefont {Mintert},\ and\ \citenamefont {Buchleitner}}]{tichyPRA}%
  \BibitemOpen
  \bibfield  {author} {\bibinfo {author} {\bibfnamefont {M.~C.}\ \bibnamefont
  {Tichy}}, \bibinfo {author} {\bibfnamefont {F.}~\bibnamefont {Mintert}}, \
  and\ \bibinfo {author} {\bibfnamefont {A.}~\bibnamefont {Buchleitner}},\
  }\href@noop {} {\bibfield  {journal} {\bibinfo  {journal} {Phys. Rev. A}\
  }\textbf {\bibinfo {volume} {87}},\ \bibinfo {pages} {022319} (\bibinfo
  {year} {2013})}\BibitemShut {NoStop}%
\bibitem [{\citenamefont {Killoran}\ \emph {et~al.}(2014)\citenamefont
  {Killoran}, \citenamefont {Cramer},\ and\ \citenamefont
  {Plenio}}]{plenio2014PRL}%
  \BibitemOpen
  \bibfield  {author} {\bibinfo {author} {\bibfnamefont {N.}~\bibnamefont
  {Killoran}}, \bibinfo {author} {\bibfnamefont {M.}~\bibnamefont {Cramer}}, \
  and\ \bibinfo {author} {\bibfnamefont {M.~B.}\ \bibnamefont {Plenio}},\
  }\href@noop {} {\bibfield  {journal} {\bibinfo  {journal} {Phys. Rev. Lett.}\
  }\textbf {\bibinfo {volume} {112}},\ \bibinfo {pages} {150501} (\bibinfo
  {year} {2014})}\BibitemShut {NoStop}%
\bibitem [{\citenamefont {Peres}(1995)}]{peresbook}%
  \BibitemOpen
  \bibfield  {author} {\bibinfo {author} {\bibfnamefont {A.}~\bibnamefont
  {Peres}},\ }\href@noop {} {\emph {\bibinfo {title} {{Quantum Theory: Concepts
  and Methods}}}}\ (\bibinfo  {publisher} {Springer},\ \bibinfo {address}
  {Dordrecht, The Netherlands},\ \bibinfo {year} {1995})\BibitemShut {NoStop}%
\bibitem [{\citenamefont {Sciarrino}\ \emph {et~al.}(2011)\citenamefont
  {Sciarrino}, \citenamefont {Vallone}, \citenamefont {Cabello},\ and\
  \citenamefont {Mataloni}}]{sciarrinoPRA}%
  \BibitemOpen
  \bibfield  {author} {\bibinfo {author} {\bibfnamefont {F.}~\bibnamefont
  {Sciarrino}}, \bibinfo {author} {\bibfnamefont {G.}~\bibnamefont {Vallone}},
  \bibinfo {author} {\bibfnamefont {A.}~\bibnamefont {Cabello}}, \ and\
  \bibinfo {author} {\bibfnamefont {P.}~\bibnamefont {Mataloni}},\ }\href@noop
  {} {\bibfield  {journal} {\bibinfo  {journal} {Phys. Rev. A}\ }\textbf
  {\bibinfo {volume} {83}},\ \bibinfo {pages} {032112} (\bibinfo {year}
  {2011})}\BibitemShut {NoStop}%
\bibitem [{Note1()}]{Note1}%
  \BibitemOpen
  \bibinfo {note} {We stress that also this output state is forbidden for
  fermions, so that the protocol does not work for this kind of particles even
  if they initially come from separated spatial modes rather than from the same
  mode.}\BibitemShut {Stop}%
\bibitem [{\citenamefont {Ma}\ \emph {et~al.}(2011)\citenamefont {Ma},
  \citenamefont {Zotter}, \citenamefont {Kofler}, \citenamefont {Jennewein},\
  and\ \citenamefont {Zeilinger}}]{zeilingerPRA}%
  \BibitemOpen
  \bibfield  {author} {\bibinfo {author} {\bibfnamefont {X.-S.}\ \bibnamefont
  {Ma}}, \bibinfo {author} {\bibfnamefont {S.}~\bibnamefont {Zotter}}, \bibinfo
  {author} {\bibfnamefont {J.}~\bibnamefont {Kofler}}, \bibinfo {author}
  {\bibfnamefont {T.}~\bibnamefont {Jennewein}}, \ and\ \bibinfo {author}
  {\bibfnamefont {A.}~\bibnamefont {Zeilinger}},\ }\href@noop {} {\bibfield
  {journal} {\bibinfo  {journal} {Phys. Rev. A}\ }\textbf {\bibinfo {volume}
  {83}},\ \bibinfo {pages} {043814} (\bibinfo {year} {2011})}\BibitemShut
  {NoStop}%
\bibitem [{\citenamefont {Somaschi}\ \emph {et~al.}(2016)\citenamefont
  {Somaschi} \emph {et~al.}}]{somaschiNatPhoton}%
  \BibitemOpen
  \bibfield  {author} {\bibinfo {author} {\bibfnamefont {N.}~\bibnamefont
  {Somaschi}} \emph {et~al.},\ }\href@noop {} {\bibfield  {journal} {\bibinfo
  {journal} {Nat. Photon.}\ }\textbf {\bibinfo {volume} {10}},\ \bibinfo
  {pages} {340?345} (\bibinfo {year} {2016})}\BibitemShut {NoStop}%
\bibitem [{\citenamefont {Rab}\ \emph {et~al.}()\citenamefont {Rab},
  \citenamefont {Polino}, \citenamefont {Man}, \citenamefont {An},
  \citenamefont {Xia}, \citenamefont {Spagnolo}, \citenamefont {Franco},\ and\
  \citenamefont {Sciarrino}}]{sciarrino2017}%
  \BibitemOpen
  \bibfield  {author} {\bibinfo {author} {\bibfnamefont {A.~S.}\ \bibnamefont
  {Rab}}, \bibinfo {author} {\bibfnamefont {E.}~\bibnamefont {Polino}},
  \bibinfo {author} {\bibfnamefont {Z.-X.}\ \bibnamefont {Man}}, \bibinfo
  {author} {\bibfnamefont {N.~B.}\ \bibnamefont {An}}, \bibinfo {author}
  {\bibfnamefont {Y.-J.}\ \bibnamefont {Xia}}, \bibinfo {author} {\bibfnamefont
  {N.}~\bibnamefont {Spagnolo}}, \bibinfo {author} {\bibfnamefont {R.~L.}\
  \bibnamefont {Franco}}, \ and\ \bibinfo {author} {\bibfnamefont
  {F.}~\bibnamefont {Sciarrino}},\ }\href@noop {} {\bibinfo  {journal}
  {arXiv:1702.04146 [quant-ph]}\ }\BibitemShut {NoStop}%
\bibitem [{\citenamefont {Kok}\ \emph {et~al.}(2002)\citenamefont {Kok},
  \citenamefont {Lee},\ and\ \citenamefont {Dowling}}]{QND1}%
  \BibitemOpen
\bibfield  {journal} {  }\bibfield  {author} {\bibinfo {author} {\bibfnamefont
  {P.}~\bibnamefont {Kok}}, \bibinfo {author} {\bibfnamefont {H.}~\bibnamefont
  {Lee}}, \ and\ \bibinfo {author} {\bibfnamefont {J.~P.}\ \bibnamefont
  {Dowling}},\ }\href@noop {} {\bibfield  {journal} {\bibinfo  {journal} {Phys.
  Rev. A}\ }\textbf {\bibinfo {volume} {66}},\ \bibinfo {pages} {063814}
  (\bibinfo {year} {2002})}\BibitemShut {NoStop}%
\bibitem [{\citenamefont {Munro}\ \emph {et~al.}(2005)\citenamefont {Munro},
  \citenamefont {Nemoto}, \citenamefont {Beausoleil},\ and\ \citenamefont
  {Spiller}}]{QND3}%
  \BibitemOpen
  \bibfield  {author} {\bibinfo {author} {\bibfnamefont {W.~J.}\ \bibnamefont
  {Munro}}, \bibinfo {author} {\bibfnamefont {K.}~\bibnamefont {Nemoto}},
  \bibinfo {author} {\bibfnamefont {R.~G.}\ \bibnamefont {Beausoleil}}, \ and\
  \bibinfo {author} {\bibfnamefont {T.~P.}\ \bibnamefont {Spiller}},\
  }\href@noop {} {\bibfield  {journal} {\bibinfo  {journal} {Phys. Rev. A}\
  }\textbf {\bibinfo {volume} {71}},\ \bibinfo {pages} {033819} (\bibinfo
  {year} {2005})}\BibitemShut {NoStop}%
\bibitem [{\citenamefont {Sathyamoorthy}\ \emph {et~al.}(2014)\citenamefont
  {Sathyamoorthy}, \citenamefont {Tornberg}, \citenamefont {Kockum},
  \citenamefont {Baragiola}, \citenamefont {Combes}, \citenamefont {Wilson},
  \citenamefont {Stace},\ and\ \citenamefont {Johansson}}]{QND4}%
  \BibitemOpen
  \bibfield  {author} {\bibinfo {author} {\bibfnamefont {S.~R.}\ \bibnamefont
  {Sathyamoorthy}}, \bibinfo {author} {\bibfnamefont {L.}~\bibnamefont
  {Tornberg}}, \bibinfo {author} {\bibfnamefont {A.~F.}\ \bibnamefont
  {Kockum}}, \bibinfo {author} {\bibfnamefont {B.~Q.}\ \bibnamefont
  {Baragiola}}, \bibinfo {author} {\bibfnamefont {J.}~\bibnamefont {Combes}},
  \bibinfo {author} {\bibfnamefont {C.~M.}\ \bibnamefont {Wilson}}, \bibinfo
  {author} {\bibfnamefont {T.~M.}\ \bibnamefont {Stace}}, \ and\ \bibinfo
  {author} {\bibfnamefont {G.}~\bibnamefont {Johansson}},\ }\href@noop {}
  {\bibfield  {journal} {\bibinfo  {journal} {Phys. Rev. Lett.}\ }\textbf
  {\bibinfo {volume} {112}},\ \bibinfo {pages} {093601} (\bibinfo {year}
  {2014})}\BibitemShut {NoStop}%
\bibitem [{\citenamefont {Ra}\ \emph {et~al.}(2013)\citenamefont {Ra},
  \citenamefont {Tichy}, \citenamefont {Lim}, \citenamefont {Kwon},
  \citenamefont {Mintert}, \citenamefont {Buchleitner},\ and\ \citenamefont
  {Kim}}]{Ra2013}%
  \BibitemOpen
  \bibfield  {author} {\bibinfo {author} {\bibfnamefont {Y.-S.}\ \bibnamefont
  {Ra}}, \bibinfo {author} {\bibfnamefont {M.~C.}\ \bibnamefont {Tichy}},
  \bibinfo {author} {\bibfnamefont {H.-T.}\ \bibnamefont {Lim}}, \bibinfo
  {author} {\bibfnamefont {O.}~\bibnamefont {Kwon}}, \bibinfo {author}
  {\bibfnamefont {F.}~\bibnamefont {Mintert}}, \bibinfo {author} {\bibfnamefont
  {A.}~\bibnamefont {Buchleitner}}, \ and\ \bibinfo {author} {\bibfnamefont
  {Y.-H.}\ \bibnamefont {Kim}},\ }\href@noop {} {\bibfield  {journal} {\bibinfo
   {journal} {PNAS}\ }\textbf {\bibinfo {volume} {110}},\ \bibinfo {pages}
  {1227} (\bibinfo {year} {2013})}\BibitemShut {NoStop}%
\bibitem [{\citenamefont {Chirolli}\ \emph {et~al.}(2010)\citenamefont
  {Chirolli}, \citenamefont {Burkard}, \citenamefont {Kumar},\ and\
  \citenamefont {DiVincenzo}}]{PhysRevLett.104.230502}%
  \BibitemOpen
  \bibfield  {author} {\bibinfo {author} {\bibfnamefont {L.}~\bibnamefont
  {Chirolli}}, \bibinfo {author} {\bibfnamefont {G.}~\bibnamefont {Burkard}},
  \bibinfo {author} {\bibfnamefont {S.}~\bibnamefont {Kumar}}, \ and\ \bibinfo
  {author} {\bibfnamefont {D.~P.}\ \bibnamefont {DiVincenzo}},\ }\href@noop {}
  {\bibfield  {journal} {\bibinfo  {journal} {Phys. Rev. Lett.}\ }\textbf
  {\bibinfo {volume} {104}},\ \bibinfo {pages} {230502} (\bibinfo {year}
  {2010})}\BibitemShut {NoStop}%
\bibitem [{\citenamefont {Johnson}\ \emph {et~al.}(2010)\citenamefont {Johnson}
  \emph {et~al.}}]{QND2}%
  \BibitemOpen
  \bibfield  {author} {\bibinfo {author} {\bibfnamefont {B.~R.}\ \bibnamefont
  {Johnson}} \emph {et~al.},\ }\href@noop {} {\bibfield  {journal} {\bibinfo
  {journal} {Nat. Phys.}\ }\textbf {\bibinfo {volume} {6}},\ \bibinfo {pages}
  {663} (\bibinfo {year} {2010})}\BibitemShut {NoStop}%
\bibitem [{\citenamefont {Romero}\ \emph {et~al.}()\citenamefont {Romero},
  \citenamefont {Solano},\ and\ \citenamefont {Lamata}}]{solano2017}%
  \BibitemOpen
  \bibfield  {author} {\bibinfo {author} {\bibfnamefont {G.}~\bibnamefont
  {Romero}}, \bibinfo {author} {\bibfnamefont {E.}~\bibnamefont {Solano}}, \
  and\ \bibinfo {author} {\bibfnamefont {L.}~\bibnamefont {Lamata}},\
  }\href@noop {} {\bibinfo  {journal} {arXiv:1606.01755 [quant-ph]}\
  }\BibitemShut {NoStop}%
\bibitem [{\citenamefont {Notermans}\ \emph {et~al.}(2016)\citenamefont
  {Notermans}, \citenamefont {Rengelink},\ and\ \citenamefont
  {Vassen}}]{PhysRevLett.117.213001}%
  \BibitemOpen
\bibfield  {journal} {  }\bibfield  {author} {\bibinfo {author} {\bibfnamefont
  {R.~P. M. J.~W.}\ \bibnamefont {Notermans}}, \bibinfo {author} {\bibfnamefont
  {R.~J.}\ \bibnamefont {Rengelink}}, \ and\ \bibinfo {author} {\bibfnamefont
  {W.}~\bibnamefont {Vassen}},\ }\href@noop {} {\bibfield  {journal} {\bibinfo
  {journal} {Phys. Rev. Lett.}\ }\textbf {\bibinfo {volume} {117}},\ \bibinfo
  {pages} {213001} (\bibinfo {year} {2016})}\BibitemShut {NoStop}%
  \bibitem [{\citenamefont {Wang}\ \emph {et~al.}(2016)\citenamefont {Wang} \emph
  {et~al.}}]{exptenphoton}%
  \BibitemOpen
  \bibfield  {author} {\bibinfo {author} {\bibfnamefont {X.-L.}\ \bibnamefont
  {Wang}} \emph {et~al.},\ }\href@noop {} {\bibfield  {journal} {\bibinfo
  {journal} {Phys. Rev. Lett.}\ }\textbf {\bibinfo {volume} {117}},\ \bibinfo
  {pages} {210502} (\bibinfo {year} {2016})}\BibitemShut {NoStop}%
\end{thebibliography}

%

\end{document}